\documentstyle[12pt]{article}

\renewcommand{\baselinestretch}{1.1}

\let \nn = \nonumber

\let\mx=\makebox
\def\beq{\begin{equation}}
\def\eeq{\end{equation}}
\def\bea{\begin{eqnarray}}
\def\eea{\end{eqnarray}}

\def\co#1{{\ifmmode{\cal O}_{#1}\else${\cal O}_{#1}$\fi}}
\def\fr#1.#2.{{#1\over #2}}
\let\FR =\fr
\let\CR =\cr
\def\cs#1{{\ifmmode{\cal S}_{#1}\else${\cal S}_{#1}$\fi}}
\def\at{{\ifmmode{\tilde A}\else$\tilde A$\fi}}

\begin{document}

\begin{titlepage}

\hspace*{4.45in}    OHSTPY-HEP-T-96-026\\
\hspace*{5.15in}    CERN-TH/96-316\\
\hspace*{5.50in}    DESY 96-226\\

\vskip 0.5in

\begin{center}

{\large \bf 
A Global $\chi^2$ Analysis of Electroweak Data in SO(10) SUSY GUTs}

\vskip 0.4in

Tom\'{a}\v{s} Bla\v{z}ek$^{a*}$,
Marcela Carena$^{b,c}$, 
Stuart Raby$^{a\dagger}$ and 
Carlos E.M. Wagner$^{b}$ \\ 

\vskip 0.2in

$^a${\em Department of Physics, The Ohio State University,}\\
    {\em 174 W. 18th Ave., Columbus, OH 43210} \\
\vskip 0.1in
$^b${\em Theory Division, CERN, Geneva, Switzerland}\\
\vskip 0.1in
$^c${\em Deutsches Elektronen-Synchrotron, Hamburg, Germany}\\
\vskip 0.2in

November 1, 1996 \\

\end{center}

\vskip 0.7in

\begin{abstract}
We present the details of a global $\chi^2$ analysis of 
electroweak data, including fermion masses and mixing angles, in SO(10) 
SUSY GUTs.  Just as precision electroweak data is used to test the Standard 
Model, the well determined Standard Model parameters are the precision
electroweak data for testing theories beyond the Standard Model. In this 
paper we use the latest experimentally measured values for these parameters. 
We study several models discussed in the literature. One of these models 
provides an excellent fit to the low energy data with $\chi^2 \sim 1$ for 
3 degrees of 
freedom.  We present graphs of constant $\chi^2$ contours as functions of 
position in soft SUSY breaking parameter space, as well as our predictions for 
a few selected points in parameter space.  We also study the sensitivity of our
results to changes in various parameters.  Finally, we discuss the 
consequences of our work in the context of a general MSSM analysis at the Z 
scale.

\end{abstract}

\vskip 0.5in


\vskip 0.1in

$^*${\footnotesize On leave of absence from 
the Dept. of Theoretical Physics, Comenius Univ., Bratislava, Slovakia;\\ 
\makebox[2.1em]{ }blazek@mps.ohio-state.edu} \\

$^\dagger${\footnotesize raby@mps.ohio-state.edu}

\end{titlepage}


\renewcommand{\thepage}{\arabic{page}}
\setcounter{page}{1}
\renewcommand{\baselinestretch}{1.3}

\section{Introduction}

The Standard Model is by far one of the greatest achievements of 
the twentieth century.  Its stature is further elevated by the 
fact that after a decade or more of testing there is still no 
evidence for physics beyond the Standard Model(SM).  Yet we 
nevertheless remain firm in our belief that the SM cannot be the
whole story.  For one thing there are 18 phenomenological 
parameters which
await derivation from a more fundamental theory.  In particular 13 of these
parameters are related to the fermion mass sector.  Secondly, there
is growing evidence that neutrinos may have mass; yet another indication of
physics beyond the SM.  Finally, a quantum theory of gravity has yet to be
incorporated into the SM.  Clearly string theory is the leading candidate for a
fundamental theory of Nature;  it can incorporate gravity, as well as the
known quarks, leptons and gauge interactions.  The problem lies in finding the
matching conditions between this fundamental string theory and the effective
field theory which describes the physics below the Planck (or string)
scale.\footnote{Within string theory the problem is different; it is 
to understand string dynamics and why one particular vacuum is preferred over 
all others.}

 The fermion mass sector is now well measured.  It is clear that fermions come
 in well defined families with an amazingly simple regularity in the hierarchy
of masses and mixing angles.  It is hoped that by understanding this sector we
 can find the correct matching conditions at the string scale.  Now seems an
 appropriate time to begin such an endeavor. It was only ten years ago that
 the top quark mass and the CKM angle $V_{ub}$ were unknown; while 
 $V_{cb}$ and the weak mixing angle $sin^2\theta_W$ were measured, but with 
 large error bars.  Today, as a result of both experimental and theoretical 
 progress, all but one of the 18 parameters of the Standard Model are known to 
 much better accuracy [the Higgs mass has yet to be measured].  These 
 parameters are the precision 
electroweak data for testing theories of fermion masses.
 
Of course once setting out to test theories of fermion masses,
we must necessarily choose some particular theories to test.  It is in this
choice of a theory that we now invoke some theoretical prejudices.
It seems clear to us that the simple observed pattern of 
fermion masses and mixing angles is not due to some {\em random} dynamics at 
an effective cut-off scale $M$ but is instead evidence that a small set of 
fermion mass operators are dominant.  Thus we are lead to postulate
 --  \begin{enumerate} \item  that a few effective operators at an 
effective cut-off scale $M$ (where $M = M_{Planck} \;{\rm or}\; M_{string}$) 
(or at a GUT scale $M_G$) dominate the quantitative behavior of fermion masses 
and mixing angles; and  \item  that a more fundamental theory, incorporating 
Planck or stringy dynamics, will generate an effective
field theory below $M$ including these dominant operators.  \end{enumerate}  

We are thus encouraged to find this effective field 
theory, thereby determining the matching conditions near the Planck scale. 
 However, in order to make progress, it is clear that {\em order of 
magnitude} comparisons with the data are insufficient.   Moreover, since the 
predictions of any simple theory are correlated, a {\em significant test} of 
any theory requires a {\em global fit to all the low energy data}.  This paper 
makes the first attempt to bring the tests of theories of fermion
masses into conformity with the accuracy of the low energy data.

In this paper we present a global $\chi^2$ analysis of precision electroweak 
data, {\em including fermion masses and mixing angles}, within the context of 
several theories of fermion masses based on SO(10) SUSY GUTs.  Of course, the 
self-consistent global analysis we describe can be applied to any predictive 
theory.  

Why look at SO(10) SUSY GUTs?
\begin{itemize}
\item We use SUSY GUTs since they give the simplest explanation for the 
experimental observation that the three gauge couplings appear to meet at a 
scale of order $10^{16}$ GeV\cite{DRW}.   \item We use SO(10) since it 
provides the simplest explanation for the observed family structure of the 
light fermions\cite{SO(10)}.  \end{itemize}

An important feature of our analysis is the inclusion of low energy 
supersymmetric threshold corrections.  We study their significance for 
minimizing $\chi^2$. As has been shown earlier \cite{tanbetcorr}\cite{bpr}, 
with  tan$\beta$ large there are potentially large corrections to those 
observables which are tan$\beta$ suppressed  at tree level. These observables 
include, for instance, masses of $b,\,s$ and $d$ quarks,  and masses of charged 
leptons.  The CKM elements $V_{cb}$ and $V_{ub}$, resulting from the mismatch
between up- and down-type quark mass matrices, also get these corrections.  In
this case the large corrections to the $d$-quark mass matrix are passed on to the
unitary  matrices which diagonalize it \cite{bpr}.  
All these corrections could potentially conspire and open a window in 
parameter space which has been unnoticed in the analysis neglecting them. 
In fact, such interplay is hard to study in the usual bottom-up scenario, but 
simple to implement in a global top-down analysis like this one.

In section 2 we describe the low energy observables which enter the $\chi^2$ 
analysis,  their measured values and the associated experimental and/or 
theoretical errors.  

In section 3 we discuss the self-consistent evaluation of the theoretical 
predictions, starting with the effective SO(10) GUT and the resulting boundary 
conditions at $M_G$; then detailing the renormalization group running to $M_Z$ 
and the threshold corrections included at the weak scale.  

In section 4 we discuss models with four effective mass operators
at the GUT scale.  We show in detail results for the best
model out of nine ADHRS\cite{adhrs} models.  In this model [model 4]
the number of arbitrary Yukawa parameters $n_y = 5$.  The best fits give 
$\chi^2 \sim 13-14$ for 5 degrees of freedom.  All low energy observables are fit 
to within their $2 \sigma$ bounds.  The fits for 4 observables, however, 
($V_{cb}, \; V_{ub}/V_{cb}$, the Kaplan-Manohar-Leutwyler ellipse parameter $Q$ 
and the Jarlskog CP violating invariant $J$) lie close to the $2 \sigma$ border.
We show that these results are true predictions of the theory, i.e. we see no 
other way to improve the results, {\bf except} with the addition of
a new operator contributing 
to the 13 and 31 elements of the Yukawa matrices \footnote{Note that models
with smaller values of $\tan\beta$ not only involve  an extra
free parameter but also require additional operators to correctly
reproduce the experimental values for the 
first generation quark masses \cite{CDRW}.}.

In section 5 we analyze 3 models derived from complete SUSY GUTs discussed 
recently by Lucas and S.R.\cite{lucas}.  One of these three models is identical 
to model 4 of ADHRS when states with mass greater than $M_G$ are integrated out;
this is model 4b.  The other two include one new effective operator at the GUT 
scale which distinguishes the two models, 4a and 4c.  We find that model 4a 
gives a best fit with $\chi^2 \sim 3$ for 3 dof in a small corner of parameter
space, while model 4c gives 
$\chi^2 \leq 1$ for 3 dof over a large subspace of the allowed parameter space.   

Our main results are found in figures [1 - 4] and tables (III - IX).  These 
include graphs of constant $\chi^2$ contours in the  two dimensional soft SUSY 
parameter space labelled by $m_0$, a universal squark and slepton mass 
parameter, and $M_{1/2}$, a universal gaugino mass parameter both with values 
given at $M_G$.   We have several such plots for different values of the Higgs 
mass parameter $\mu$, given at $M_Z$.  In the tables we give the computed
values for the low energy observables for a few selected points in parameter 
space. We also include values for the Higgs and SUSY spectra, the CP violating angles 
$\alpha,\; \beta$ and $\gamma$, measurable in neutral B decays, the Wolfenstein
parameters $\eta, \; \rho$, and the ratio $m_u/m_d$ for these  points.  Note, in
figure 5, we present a scatter plot for values of ($\sin 2\alpha, \; \sin 2\beta$) 
predicted by models 4a and 4c for points in figures 3(a - c) and 
4(a - c), respectively, with $\chi^2$ values $\le 3$.  Finally we test the sensitivity 
of $\chi^2$ to changes in various parameters.

Section 6 contains further discussion of our results and conclusions.

\section{Low energy observables -- experimental values}

Our $\chi^2$ function includes 20 low energy observables.  In addition we have 
incorporated the experimental bounds on sparticle masses into the code as a  
penalty in $\chi^2$; added if one of these bounds is violated.  This gaurantees 
that we remain in the experimentally allowed regions of parameter space. Let 
us now discuss the experimental observables and their errors.  

We have 
\begin{itemize}

\item 6 parameters associated with the Standard Model gauge and electroweak 
symmetry breaking sectors --- $ \alpha_s(M_Z), \; \alpha_{EM},  \; G_{\mu} ,
\;M_W,\;  M_Z, \;\; {\rm and}\;  \rho_{new}$; 
\item 13 parameters associated with fermion masses and
mixing angles --- $
M_t, \; m_b(M_b), \; (M_b-M_c), \; m_s, \; m_d/m_s, \; Q^{-2}, \;
M_{\tau}, \; M_{\mu}, \; M_e, \; V_{us}, \; V_{cb}, \; V_{ub}/V_{cb},\; 
{\rm and} \;  J~;\footnote{In the actual $\chi^2$ analysis $J$ is replaced by 
the hadronic matrix element $\hat B_K$ as will be explained later in this section.}$
 and 
\item the branching ratio  for $b \rightarrow s \gamma$,
\end{itemize}
where $\rho_{new}$ includes the added contribution to the electroweak $\rho$ 
parameter from new physics beyond the SM, $Q$ is the Kaplan-Manohar-Leutwyler 
ellipse parameter\cite{kml} relating $u, \,d,$ and $s$ quark masses and $J$ is 
the Jarlskog CP violating invariant\cite{jarlskog} (see more on these 
parameters below). Note also, that mass parameters denoted with a capital 
letter $M$ are defined as pole masses,  while $m_b, \;m_s,\; m_d,\; m_u$ are 
defined as the $\overline{MS}$ running  masses. 
The three light quark running masses are evaluated at the scale 1 GeV. 

The experimental values for the observables are given in table 
\ref{t:observables} with the associated experimental or theoretical error 
$\sigma$.  {\em Note, $\sigma$ is taken to be either the experimental error or 
1/2\%, whichever is larger.}  This is because our theoretical calculation also
introduces an error; the numerical solution to the renormalization group 
equations is computed with some limited precision and the values of all 
observables are calculated within the first orders of the perturbative 
expansion in couplings. Thus 1/2\% represents a conservative simple estimate 
of the combined theoretical error assigned to each low energy theoretical output
\footnote{For $G_{\mu}$ we take $\sigma$ to be  1\%.  This is to account for 
the additional error resulting from our neglecting one loop SUSY box 
and vertex corrections.}. 
As a consequence, the listed errors for $M_Z,\; M_W,\; G_{\mu},\; \alpha_{EM},
\; M_{\tau},\; M_{\mu},\; M_e$ are dominated by theoretical uncertainties.

For $\alpha_s(M_Z)$ we use the central value quoted by Schmelling in
Warsaw\cite{asW}. However, for the error we choose a conservative value as
suggested by Webber\cite{glasgow} or Burrows\cite{asC}.   This accounts for the
fact that the systematic errors of individual $\alpha_s$ measurements are
 significant and thus caution is required when adding them together.

Note that the traditional Standard Model parameter  
$\sin^2\theta_W(M_Z)|_{\overline{MS}}\;$  
is not among the observables in table \ref{t:observables}. Neither is $J$, 
$\epsilon_K$ or any other standard CP violating quantity listed. Instead, in the
 list of observables, $\sin^2\theta_W(M_Z)$ is replaced by the Fermi constant 
$G_{\mu}$, and $J$ is replaced by the bag constant $\hat B_K$.  
$G_{\mu}$ is substituted for $\sin^2\theta_W(M_Z)$  since the former is 
extracted from the muon lifetime formula \cite{mu} with very little sensitivity 
to the particle content at the scale $M_Z$ and the underlying theory at this 
scale.  In fact, it could be defined completely in the context of the effective 
four-fermi contact interaction plus electromagnetism, and the remaining 
uncertainty of the order $M_{\mu}^2/M_W^2$ would then appear in the quantities 
derived from $G_{\mu}$.  On the contrary, the experimental value of 
$\sin^2\theta_W(M_Z)$ in the $\overline{MS}$ scheme is derived 
from the precision measurements at the Z peak. Loop effects play an important 
role in its precise determination and hence one has to specify the theory at 
$M_Z$ and its particle content, and then calculate all relevant radiative 
corrections. Thus the underlying theory is an explicit input for the extraction 
of the experimental value of $\sin^2\theta_W(M_Z)$ and, naturally, it is 
commonly taken to be the Standard Model (SM).  However, we adopt an approach 
where the minimal supersymmetric extension of the Standard Model (MSSM) is 
matched directly to the low energy $SU(3)_c \times U(1)_{em}$ effective theory 
by integrating out the superpartners at the same time as the top quark and W/Z 
bosons (at a single scale $M_Z$); hence there is no room for the SM.   This 
approach is particularly sensible if there are light SUSY particles (below $M_Z$), 
such as charginos, neutralinos, pseudoscalar Higgs, etc., for then there is no 
energy regime where the SM dominates\cite{sin2thetaw}.
Although not among the observables contributing to the $\chi^2$ function, 
the SM value of $\sin^2\theta_W(M_Z)$ is displayed among the model predictions 
in tables \ref{t:model4cpred} and \ref{t:model4cpredIII}.

The CP violating parameter $J$ is defined by the expression -- \beq J =
Im(V_{ud}V_{ub}^*V_{cb}V_{cd}^*) \approx |V_{cd}| |V_{ub}/V_{cb}| |V_{cb}|^2 \;\sin{\xi}
\label{eq:J}  \eeq where $\xi$ is the CP violating phase.
We test $J$ by a comparison to the experimental value extracted from 
the well-known $K^0-\overline{K^0}$ mixing observable 
$\epsilon_K =  (2.26 \pm 0.02)\times10^{-3}$.  The largest uncertainty in such
a comparison, however, comes in the value of the QCD bag constant $\hat B_K$. 
 We thus exchange the Jarlskog parameter $J$ for $\hat B_K$ in the list 
of low-energy data we are fitting. Our theoretical value of $\hat B_K$ is 
defined as that value needed to agree with $\epsilon_K$
for a set of fermion masses and mixing angles derived from the
GUT-scale. We test this theoretical value against the ``experimental'' 
value of $\hat B_K$, which is given in table \ref{t:observables}. This value, 
together with its error estimate, is obtained from recent lattice 
calculations\cite{kilcup}.

The experimental value for $\rho_{new}$ is obtained from Langacker's combined 
fits to the precision electroweak data, presented at recent 
workshops\cite{langacker}.

For fermion masses and mixing angles we use those combinations of parameters 
which are known to have the least theoretical and/or experimental uncertainties.
For example, while the bottom and top quark masses are known reasonably well, 
the charm quark mass is not known as accurately.  On the other hand, heavy 
quark effective theory relates the mass difference $M_b - M_c$ between the 
bottom and charm quark {\em pole masses} to about 5\% accuracy\cite{hqet}.  We thus 
use this relation, instead of the charm quark mass itself, to
test the theory.  We note that $M_b$ and $M_c$ are calculated from the 
$\overline{MS}$ running masses, $m_b(M_b)$ and $m_c(M_c)$ using two loop QCD 
threshold corrections.  In fact there are infra-red renormalons which make this
perturbative evaluation ambiguous.  However, these renormalon contributions cancel
in the mass difference.  Moreover, it has been argued that this ambiguity cancels
when extracting 
$V_{cb}$ via semileptonic B decays\cite{hqet}\cite{bbb}.\footnote{The correction to
the bottom mass $(M_b - m_b(M_b))/m_b(M_b)$ {\it at two loops} is $\approx 25$\%
less than the results found recently by Ball, Beneke and Braun using an infinite
order resummation within a naive non-abelianization of the mass
correction\cite{bbb}.  The theoretical error for this result is also estimated to
be about 25\%.  By using two loop pole masses, combined with the mass relation for
$M_b - M_c$, we obtain a small value for $m_c(M_c)$.  This value increases when
using the  scheme of BBB\cite{bbb}.}

Similarly, among the three light quarks there is one good relation, 
(reparametrization invariant and free of $O(m_q)$ corrections) which severely 
constrains any theory of fermion masses.   This is the Kaplan-Manohar-Leutwyler 
ellipse given by 
\beq  1 = {1 \over Q^2} {m_s^2 \over m_d^2} + {m_u^2 \over m_d^2}  
\eeq
or 
\beq  Q = {{m_s \over m_d} \over \sqrt{1 - {m_u^2 \over m_d^2}}} \eeq
where  $Q$ is the ellipse parameter.  The experimental value for $Q$ is 
obtained from a weighted average of lattice results and a chiral Lagrangian 
analysis, with significant contributions from the violation of Dashen's 
theorem\cite{leutwyler}. We follow Donoghue\cite{donoghue} and assign a 
conservative 10\% error to 
the experimental value of $Q^{-2}$.  Note that $m_d/m_s$ derived from chiral 
Lagrangian analysis is not free of first order quark mass corrections, and 
hence $\sigma$ is much smaller for $Q^{-2}$ than for 
$m_d/m_s$\cite{donoghue}.  In addition, we do not constrain $m_u/m_d$
independently, as suggested by Leutwyler\cite{leut}, which requires additional
input from an expansion in $1/N_{color}$.  Instead we quote values for $m_u/m_d$
as output.  Finally we use $m_s$ given by the PDG\cite{pdg}.  At the end of
section V we consider the lower values for $m_s$ suggested by recent lattice
calculations\cite{lattice:ms}.

The remaining parameters are more or less self evident.  We just remark that 
the central value for $V_{cb}$, as well as the error bars, has steadily 
decreased in the last 5 years, making it a very significant
constraint.\footnote{Note, the actual value of $V_{cb}$ is correlated with the 
values of $m_b$ and 
$m_c$\cite{hqet}.  We have not included this correlation in our analysis.}
In addition, the value for $V_{ub}/V_{cb}$ changed dramatically in 1992. It 
changed from approximately $0.15 \pm 0.05$ to its present value $0.08 \pm 0.02$, 
where the errors were and continue to be dominated by theoretical
model dependence.  Clearly the systematic uncertainties were large but are now 
hopefully under control.

\section{Low energy observables -- computed values}

In our analysis we consider the minimal supersymmetric standard model defined 
at a GUT scale $M_G$ with, in all cases  (but one), tree level GUT boundary 
conditions on gauge couplings and Yukawa matrices.  In this one case, we 
include an arbitrary parameter, $\epsilon_3$, given by
\begin{equation}
\epsilon_3 \equiv {\alpha_3(M_G) - \tilde{\alpha}_G \over \tilde{\alpha}_G}
\;\; {\rm with} \;\;  \tilde{\alpha}_G \equiv \alpha_1(M_G) = \alpha_2(M_G)
\label{eq:eps3} \end{equation}
 which parametrizes the one
loop  threshold correction to gauge coupling unification.\footnote{$\epsilon_3$ is 
calculable in any complete SUSY GUT.  It is also constrained somewhat by the 
bounds on the nucleon lifetime\cite{lucas}.}  We also include 7 soft SUSY 
breaking parameters ---   an overall scalar mass $m_0$ for squarks and
sleptons, a common gaugino mass $M_{1/2}$, the parameters $A_0$, $B$ and 
$\mu$, and in addition we have allowed for non-universal Higgs masses, 
$m_{H_u}$ and $m_{H_d}$.\footnote{Note that if the messenger scale of SUSY 
breaking is $M_{Planck}$ then our analysis is not completely self-consistent. 
In any complete SUSY GUT defined up to an effective cut-off scale $M > M_G$, 
the interactions above $M_G$ will renormalize the soft breaking parameters.  
This will, in general, split the degeneracy of squark and slepton masses at 
$M_G$ even if they are degenerate at $M$.  On the other hand, bounds on flavor
changing neutral current processes, severely constrain the magnitude of 
possible splitting. Thus these corrections must be small. In addition, in 
theories where SUSY breaking is mediated by gauge exchanges  with a messenger 
scale below (but near) $M_G$,  the present analysis is expected to apply 
unchanged, since in this case squarks and sleptons will be nearly degenerate 
at the messenger scale.  The Higgs mases, on the other hand, are probably
dominated by new interactions which also generate a $\mu$ term.  It is thus 
plausible to expect the Higgs masses to be split and independent of squark and 
slepton masses.  The parameter $A_0$ could also be universal at the messenger 
scale.}   Thus the number of arbitrary parameters in the effective GUT includes 
the 3 gauge parameters, 7 soft SUSY breaking parameters and $n_y$ Yukawa 
parameters. The number $n_y$ and the form of the Yukawa matrices are model 
dependent.   The models are discussed in sections IV and V.

The effective theory between $M_G$ and $M_Z$ is the MSSM. We use two loop SUSY 
renormalization group equations [RGE] for dimensionless parameters and 
one loop RGE for dimensionful parameters from $M_G$ to $M_Z$. However we have 
checked that the corrections to our results obtained by using two loop RGE for 
dimensionful parameters\cite{mv} are insignificant. When in the framework of 
the MSSM we use Dimensional Reduction regularization in the Modified Minimal 
Subtraction Scheme. This renormalization  procedure will be abbreviated as 
$\overline{DR}$ in the following text. 

When crossing the $M_Z$ scale we match the MSSM to the non-supersymmetric 
$SU(3)_c\times U(1)_{em}$ gauge theory, i.e.\ our effective theory below  the 
$Z$  threshold is standard QCD and electromagnetism. In this effective theory 
we perform the calculations using Dimensional Regularization in the Modified 
Minimal Subtraction scheme. Quantities renormalized according to this 
prescription will appear with a standard subscript $\overline{MS}$ in the 
following formulas. Below the $Z$ scale, three loop QCD renormalization group 
equations run quark masses and $\alpha_s$ down to lower scales. When this 
running crosses the $b$ and $c$ mass thresholds the number of flavors 
is reduced by one each time. Finally we end up with the three flavor QCD RGE 
as we arrive at the scale of $1\,GeV$, where the three light quark masses are 
evaluated.  In addition, QED running is included to one loop precision. 
(For further details see \cite{ramond}.)

At $M_Z$ we calculate complete one loop MSSM 
corrections to the W and Z masses.  Thus  the W and Z pole masses are given by 
the formulae:
\bea  M_W^2 \!\! &=&\! {1 \over 4} \,g_2^2 \, v^2 +  \delta M_W^2,  \\ 
      M_Z^2 \!\! &=&\! {1 \over 4}\,({3 \over 5}\,g_1^2 + g_2^2) \, v^2 + 
      \delta M_Z^2, 
\eea where $\delta M_W^2$ and $\delta M_Z^2$ are the complete one loop 
self-energies \cite{sin2thetaw},\cite{hollik} 
in the $\overline{DR}$ scheme and $g_1,\, g_2$ are the 
$\overline{DR}$ gauge couplings evaluated at $M_Z$ in the MSSM.  

The Higgs vacuum expectation value $v$ is an implicit function of soft SUSY 
breaking parameters and gauge and Yukawa couplings.   It is determined by self 
consistently demanding minimization of the tree level Higgs potential.  
The actual value for $v$ is found by minimizing $\chi^2$.

The theoretical value of $G_{\mu}$ is calculated according to the formula
  
\beq G_{\mu} = {\pi \, \alpha(M_Z) \over \sqrt{2} M_W^2\, \sin^2\!\theta_W(M_Z) }\;
                  {1 \over 1-\Delta \hat r_W^{MSSM} }  .
\label{eq:GmuTh}
\eeq
The right-hand side of this equation is consistently evaluated in the 
$\overline{DR}$ scheme within the context of the MSSM. 
The electromagnetic structure constant $\alpha$ and $\sin^2\theta_W$ are 
determined by the relations among the gauge coupling constants $\alpha_1$ and 
$\alpha_2$: 
\bea \sin^2\theta_W(M_Z)  &=&  {{3 \over 5} \alpha_1(M_Z) \over 
                                      {3 \over 5} \alpha_1(M_Z) + \alpha_2(M_Z)},
\label{eq:sw}   \\
            \alpha (M_Z)  &=&  \alpha_2(M_Z) \,\sin^2\!\theta_W(M_Z)\; .
\label{eq:alpha}
\eea 
$\Delta \hat r_W^{MSSM}$ in (\ref{eq:GmuTh}) contains the corrections from the 
$W$ self-energy and vertex and box diagrams. We follow refs. 
(\cite{sin2thetaw},\cite{sirlin}) 
closely, but not to all details: we don't include the vertex and box diagrams 
containing SUSY particles. These have a minor effect (see results in 
\cite{sin2thetaw}) and we compensate for them with an additional $1/2$\% error 
on $G_{\mu}$, as already mentioned in a previous footnote.
In fact, we use a reverse procedure to the one introduced in \cite{sin2thetaw}, 
where the equation (\ref{eq:GmuTh}) is utilized to derive the MSSM value of 
$\sin^2\theta_W$ in the bottom-up approach starting from the precise 
value of $G_{\mu}$.

When crossing the $Z$ threshold, gauge couplings are subject to two different  
kinds of corrections. That is to account for the states which are integrated out 
at this scale , and for the transition from the $\overline{DR}$ scheme to the 
$\overline{MS}$ scheme. First, when the SUSY particles, the top quark and 
the $Z$ and $W$ gauge bosons are integrated out , the threshold  corrections to 
$\alpha_s|_{\overline{DR}}\;$ and $\alpha|_{\overline{DR}}\;$ read 
\bea 
{\delta \alpha_s \over \alpha_s} = {\alpha_s \over \pi}  &(& 
   {1\over 12} 
 \sum_{i=1 \atop {(\,all\;\;\atop squarks)}}^{12}\log{m_{\tilde{q_i}}\over M_Z} + 
    \log{M_{\tilde{g}}\over M_Z} + {1\over 3}\log{M_t\over M_Z}) , 
\label{eq:das} 
\\ 
{\delta \alpha   \over \alpha  } = {\alpha  \over 2\pi}  &(& 
   {4\over 9}
\sum_{i=1 \atop {(\,{\rm u-}type\atop squarks)}}^6\log{m_{\tilde{q_i}}\over M_Z} + 
   {1\over 9}
\sum_{i=1 \atop {(\,{\rm d-}type\atop squarks)}}^6\log{m_{\tilde{q_i}}\over M_Z} + 
   {1\over 3}
\sum_{i=1 \atop {(\,charged    \atop sleptons)}}^6\log{m_{\tilde{l_i}}\over M_Z}
\nonumber \\
   \!&+&\! 
   {4\over 3}\sum_{i=1}^2\log{M_{C_i}\over M_Z} + {1\over 3}\log{M_{H^+}\over M_Z}
      + {16\over 9}\log{M_t\over M_Z} - 7\log{M_W\over M_Z} ).
\label{eq:da}
\eea 
The $\overline{DR}$ values of $\alpha_i,\; 
(i=1,2,3)$ are converted to the corresponding $\overline{MS}$ values using the 
relations 
 \beq  
{1\over \alpha_i(M_Z)|_{\overline{MS}}} = {1\over \alpha_i(M_Z)|_{\overline{DR}}} + C_i\,;  
\;\;\;\;\; {\rm with}\;\;\;  C_i = {C_2(G_i) \over 12 \pi} ,
 \label{eq:DRMS}
 \eeq  
where the quadratic Casimir $C_2(G_i) \!=\! N$ for SU(N) or equals zero for 
U(1). 

The strong coupling constant is now ready for comparison with the 
experimental value quoted in table \ref{t:observables}. 
The electromagnetic fine structure constant may stay within the 
$\overline{DR}$ scheme and it is corrected by an additional factor 0.0684 
(\cite{sin2thetaw} and the references therein) to account for the running down 
to the zero momentum transfer value as extracted from Thomson scattering.
 
Note that $\sin^2\theta_W$ in eq.(\ref{eq:sw}) represents the MSSM value, in the
$\overline{DR}$ scheme,  which can differ by a few $per\, cent$ from the SM
value - as found in 
\cite{sin2thetaw}.  For completeness, we also calculate the SM value, in 
the $\overline{MS}$ scheme, and quote both quantities in table \ref{t:model4cpred}. 
Our SM value of $\sin^2\theta_W(M_Z)$ is obtained from the SM values of $\alpha_1$ 
and $\alpha_2$ (by a relation analogous to eq.(\ref{eq:sw})). The latter are 
derived from the corresponding MSSM values by applying the threshold corrections 
similar to (\ref{eq:das}) and (\ref{eq:da}), with the $W$, $Z$ and $top$ quark terms 
absent, since these particles are kept in the SM. 
These are the leading logarithmic thresholds (LLT) which make up 
for the exclusion in the SM of the SUSY partners and non-minimal Higgs 
sector.\footnote{Explicit forms of the LLT corrections to $\alpha_1$ and
$\alpha_2$  are given e.g.\ by the equations (17) and (18) in
ref.\cite{sin2thetaw}. We  agree with the authors of \cite{sin2thetaw} that the LLT 
approximation cannot  proceed in a consistent way at this point since the
$SU(2)\times U(1)$ gauge  symmetry is broken at a scale comparable to the masses of 
sparticles being  integrated out. The difference compared to the full calculation, 
however, is  not very significant, and for our purposes will be neglected. Thus
our SM value of $\sin^2\theta_W$ contains a theoretical uncertainty of about 1\%,
(which is larger than the experimental uncertainty) similar to the theoretical
uncertainty assigned to $G_{\mu}$.}
A constant due to the change from the $\overline{DR}$ scheme to the $\overline{MS}$ 
scheme is added to $\delta \alpha_2 $, according to
eq.(\ref{eq:DRMS}).

In the absence of direct observations, the SUSY sector of the MSSM is tested 
by all observables calculated at the loop level. Obviously, the strongest 
constraints come from the phenomena which originate directly in loop effects. 
Because of this reason $\rho_{new}$, as well as the process 
$b\rightarrow s\gamma$, were added to the list of observables. Breaking of 
custodial $SU(2)$ by physics beyond the SM is tested using  
\beq  \rho_{new} = {\Pi^{WW}_{new}(0) \over M_W^2} -{\Pi^{ZZ}_{new}(0) \over M_Z^2} ,
\eeq
where $\Pi^{WW}_{new}(0)$ and $\Pi^{ZZ}_{new}(0)$ stand for the contributions 
of physics beyond the SM to the vector boson self-energies at zero momentum
\cite{langacker}.

For the fermionic sector, we have at tree level
${\bf m}_u \!=\! Y_u\, \sin\beta \,{ v \over \sqrt{2}},\; 
{\bf m}_d \!=\! Y_d\, \cos\beta \, { v \over \sqrt{2}}\,$ and ${\bf m}_e 
\!=\! Y_e\, \cos\beta \,{ v \over \sqrt{2}}\,$,  
where ${\bf m}_i$ and $Y_i$ ($i=u,d,e$) are $3\times 3$ mass and Yukawa 
matrices at $M_Z$.  In addition, at $M_Z$ we include the leading 
( $O(\tan\beta)$) one loop threshold corrections to the mass 
matrices \cite{bpr}. The one-loop corrected fermion mass matrices are then
diagonalized. The pole mass of the top quark is calculated from the diagonal 
running mass with the leading two loop QCD corrections included. 
We utilize the infrared fixed point behaviour of the top Yukawa coupling 
and the fact that $\alpha_s$ also changes only slightly between $M_{top}$ and 
$M_Z$, and evaluate the top pole mass from the relevant couplings at the 
$Z$ scale.\footnote{A more precise evaluation would use $\alpha_s(m_t)$ and 
$m_t(m_t)$ to obtain the pole mass $M_t$.  The error obtained with our 
approximation, using values at $M_Z$ instead, partially cancels the one
due to the non inclusion of the logarithmic corrections to the top mass,
leading to a value which differs by less than 2 GeV from the actual value
of $M_t$.} 
The other quark and lepton masses are run  down to their corresponding  
 mass scales as explained in the beginning of this section. 
$M_b$ and $M_c$ are  
 evaluated using the two loop QCD corrections while 
the lepton pole masses are obtained with the help of the one loop QED 
corrections. Upon diagonalizing the one loop corrected mass matrices at $M_Z$,
we calculate the complete form of the CKM mixing matrix (note that the CKM 
matrix elements do not run below the $Z$ scale), and then compute the 
theoretical value for $\hat B_K$ needed to fit $\epsilon_K$\cite{epsK}. The 
effects of SUSY box diagrams are neglected in this last calculation, while a 
potentially significant ($O(\tan\beta)$) one loop threshold correction to 
$J$ \cite{bpr} is included (automatically, since we start from the one loop 
corrected mass matrix ${\bf m}_d$)  as a leading higher order effect to the 
SM box diagram. The SUSY box diagram corrections can be as large as 15\% \cite{goto} 
and may affect the predictions for CP violating parameters, such as $\eta$ or 
$\sin 2\alpha$ and $\sin 2\beta$.  We discuss the sensitivity of our results to 
these corrections in section V.

Finally, special attention is paid to the branching ratio for 
$b \rightarrow s \gamma$.  We calculate the partial amplitudes for this 
process, as well as the amplitude for $b \rightarrow s\, gluon$ 
(required to resum leading logarithmic corrections), using the complete quark 
{\it and squark} mixing matrices at low energies \cite{bbmr}. 
Large QCD corrections are resummed to leading order following \cite{LObsg}. 
The low scale in this evaluation is set  equal to the pole mass $M_b$ of the 
$b$ quark, and the dependence of the branching ratio on this choice is 
neglected compared to the experimental error.

 We then form a $\chi^2$ function including the 20 low energy observables ---  
 6 in the gauge sector ($M_W$, $M_Z$, $\alpha_{EM}$, $G_{\mu}$, $\alpha_s$, 
 $\rho$), 13 in the fermion mass sector (9 charged fermion masses, 3 quark 
 mixing angles and $\hat{B}_K$) and the branching ratio for $b \rightarrow s \gamma$.
 This $\chi^2$ function is then minimized self-consistently using Minuit by 
 iteratively varying the GUT parameters with $m_0, \; M_{1/2}$, and $\mu$ fixed.
 The procedure involves two nested minimizations. The GUT scale parameters 
 $m_{H_d}$ and $m_{H_u}$ are varied separately in the second nested 
 minimization, for each step in the main minimization process where the rest of 
 the free parameters are varied. 
 Such a nested minimization is time consuming, but on the other 
 hand it guarantees a high precision in the optimization process which is very
 sensitive to radiative electroweak symmetry breaking. Analytic tree level 
 conditions for the potential minimum are used to determine $v$ and $\tan\beta$. 
With these technical details we conclude the discussion 
on the theoretical calculation of the twenty 
observables given in table \ref{t:observables}. 

In addition to the 20 observables discussed above, we have to deal 
with the experimental lower bounds on sparticle masses. We set lower limits of 
30 GeV on squark and slepton masses, expecting significant positive 
higher order corrections to the squared mass which would raise it above the
experimental lower bound of $45$ GeV.  We require that the lightest neutralino (if
below the LEP1 threshold)  contribute to the invisible Z partial width as measured
at LEP1 -- 
$\Gamma(Z \rightarrow\, {\rm invisibles}) = $
\mbox{$(-1.5 \pm 2.7) MeV$} \cite{invis}. 
We also demand that the lightest chargino be heavier than 65 GeV.  Finally we 
require the pseudo-scalar Higgs, A, to have mass greater than 80 GeV. Note that 
these bounds are on the tree level running masses. In table \ref{t:model4cpred} 
which shows predictions of the models discussed in the next sections, 
Higgs masses are displayed with the leading one loop corrections 
included.\footnote{This $CP$ odd state A, if light, (along with the other light
Higgs states)  receives significant corrections to its tree level mass in the regime
of large $\tan\beta$ in which we work. If the pseudoscalar mass is less than $M_Z$,
it receives negative mass squared corrections;  the same is true for the lightest
$CP$ even Higgs $h^0$ \cite{dmA}. }  As can be seen, the 80GeV lower limit on the
tree masses provides enough room  for the one loop corrected Higgs masses to stay
above the experimental limits.  In order to take all these bounds into account we
have  added a large penalty to the $\chi^2$ function whenever one of these bounds 
is exceeded, thus forcing the optimization procedure to stay within the 
experimentally allowed region in parameter space.

\section{Model 4 of ADHRS}
In this paper we have analyzed several models of fermion masses.  We have 
studied model 4 of ADHRS\cite{adhrs}.  In this model $n_y = 5$.  The model is 
defined by the following  4 operators in the effective theory at $M_G$.  

\begin{eqnarray}
\co{33} & = & 16_3 \ 10_1 \ 16_3 \\
\co{23} & = & 16_2 \ {A_2\over\at}\  10_1 \ {A_1\over\at}\  16_3 \nn\\
\co{12} & = & 16_1 \ 
\left({\at\over\cs{M}}\right)^3 \ 10_1 \ \left({\at\over\cs{M}}\right)^3\ 
16_2 \nn 
\end{eqnarray}

There are six possible choices for the 22 operator; all give the same 
$0 : 1 : 3$ Clebsch relation between up quarks, down quarks and charged 
leptons responsible for the Georgi-Jarlskog relation\cite{gj}.

\begin{eqnarray}
\co{22} = \nn\\
& (a)\;\; & 16_2 \ \fr\at.\cs{M}. \  10_1  {A_1 \over \at}  16_2 \\
& (b)\;\; &16_2 \ {\cs{G} \over \at}\  10_1 \ \fr A_1.\cs{M}. \ 16_2\nn\\
& (c)\;\; &16_2 \ \fr\at.\cs{M}. \ 10_1 \ \fr A_1.\cs{M}. \ 16_2\nn\\
& (d)\;\; &16_2 \ 10_1 \ {A_1\over \at}\  16_2\nn\\
& (e)\;\; &16_2 \ 10_1 \ \fr\at \ A_1.\cs{M}^2. \ 16_2\nn\\
& (f)\;\; &16_2 \ 10_1 \ {A_1 \cs{G}\over \at^2}\ 16_2\nn 
\end{eqnarray}

$A_1,\; A_2$ and $\tilde A$ are adjoint scalars with vacuum expectation values [vev]
in the B-L, Y or X (SU(5) invariant) directions of SO(10) and $\cs{M}$ is  a singlet with
vev of order $M$.

The resulting Yukawa matrices at $M_G$ are given by -- 

$$Y_u=\pmatrix{ 0& C & 0 \CR
C & 0 & -\FR1.3. B \CR
0 & -\FR4.3. B & A}$$
$$Y_d=\pmatrix{ 0& -27 C & 0 \CR
-27 C & E e^{i \phi} & \FR1.9. B \CR
0 & -\FR2.9. B & A}$$
$$Y_e=\pmatrix{ 0& -27 C & 0 \CR
-27 C & 3 E e^{i \phi} & B \CR
0 & 2 B & A} .$$
where the Yukawa matrices are defined by the couplings ---
$H_u Q Y_u \overline U + H_d Q Y_d \overline D + H_d L Y_e \overline E$
and, for example, the fields $Q (\overline U)$ are left-handed Weyl spinors
transforming as a weak doublet (singlet).

The best fits give $\chi^2 \sim 13-14$ for 5 degrees of freedom.  See 
figures 1a - c for contour lines of constant $\chi^2$ in
the $m_0 - M_{1/2}$ plane for different values of $\mu$.\footnote{$M_{1/2}$ 
and $m_0$ are GUT scale values, while $\mu$ is given at $M_Z$.}  Note that 
the regions with lowest $\chi^2$ contain significant (of order (4 - 9)\%) one 
loop SUSY threshold corrections to fermion masses and mixing angles.  These 
one loop SUSY threshold corrections scale roughly as 
$ {\mu M_{1/2} \over m_0^2} \;\;  {\rm or} \;\; {\mu A_{33} \over m_0^2} $.
Hence, from figures 1a - c, we see that these corrections are necessary to improve 
the agreement of the model with the data.

In table \ref{t:model4} we give the results for the point labelled ``II" in 
fig. 1a.  All low energy observables are fit to within their $2 \sigma$ 
bounds.  However the fits for 4 observables, ($V_{cb}, \; V_{ub}/V_{cb}$, 
$\hat B_K$ and $Q$) lie close to the $2 \sigma$ border.  Note that Minuit 
typically tries to equalize the contribution of all the observables  to 
$\chi^2$.  We now argue that these results are true predictions of the
theory.  

Consider the first three parameters.  It was shown by Hall and Rasin\cite{hr} 
that the relation \beq {V_{ub} \over V_{cb}} = 
\sqrt{{\lambda_u \over \lambda_c}}  \label{eq:hr}
\eeq holds for any fermion mass texture in which the 11, 13 and 31 elements 
of the mass matrices are zero and perturbative diagonalization is permitted.  
Note $\lambda_u, \; \lambda_c$ are the up and charm quark Yukawa couplings 
evaluated at a common renormalization scale.  A typical value for the 
right-hand side of the equation is  $0.05$ which is too small for the 
left-hand side by more than 20\%.

 We now show that the fits for $V_{ub}/V_{cb},\; V_{cb}$ and $\hat B_K$ are 
 correlated. Consider the formula for $\epsilon_K$ given by
\beq
\epsilon_K \approx 
\left({V_{ub} \over V_{cb}} \,V_{cb}^2 sin(\xi) \right)\;\hat B_K 
\times ({\rm one \; loop \; factors})
\eeq
where the first factor is just the Jarlskog parameter, $J$ - see 
eqn.(\ref{eq:J}).  We see that if $V_{ub}/V_{cb}$ is small, then $V_{cb}$ 
and $\hat B_K$ must be increased to compensate.  As a consequence, $V_{cb}$ 
and $\hat B_K$ are both too large.  The addition of  13 and 31 mass terms 
modifies relation (\ref{eq:hr}) and can,
in principle, accomodate larger values of $V_{ub}/V_{cb}$ and thus lower 
values for $V_{cb}$ and $\hat B_K$.

Now consider the ellipse parameter $Q$.  This parameter is strongly 
controlled by the Georgi-Jarlskog relation\cite{gj}
\beq  {m_s \over m_d} \approx {1 \over 9}{m_{\mu} \over m_e}  \eeq
which is satisfied by model 4.  This is an important zeroth order relation to 
try to satisfy.  However unless there are small calculable corrections to 
this relation, it leads to values of $m_s/m_d \sim 25$ and thus values of $Q$ 
which are too large.  Note, that introducing 13 and 31 terms in the down 
quark and charged lepton mass matrices can also, in principle, perturb the 
zeroth order Georgi-Jarlskog relation.  

Thus the disagreement between model 4 and the data seems significant.  It is 
unlikely that it can be fixed with the inclusion of small threshold corrections 
to the Yukawa relations at $M_G$.\footnote{Note, there are 2 types of threshold
corrections.  First, there are the higher dimension operator corrections to the 4
operators; obtained when integrating out the Froggatt-Nielsen massive
intermediates.  These are expected to be of order 10\%. Second, there are the
usual higher loop corrections.}  Before one adds new operators, however, it  is
worthwhile to consider the possibility that perhaps one of the experimental 
measurements is wrong.  It is interesting to ask whether the agreement with the 
data can be significantly improved by removing one contribution to the $\chi^2$ 
function; essentially discarding one piece of data\footnote{We thank M. Barnett for
bringing this idea to our attention.}.  In order to check  this possibility we have
artificially inflated the value of
$\sigma$ in
 the analysis for several observables (one at a time), to see
if this significantly improves the fit.  Our results are given in figure 2, 
where we also state, in the figure caption, the preferred value of the observable
with the inflated error.  As one can see, we find no
significant  improvements with this procedure.

On the other hand, we have found that we can indeed improve the results by 
adding one operator contributing to the 13 and 31 elements of the Yukawa 
matrices.  We discuss this possibility in the next section.  The additional 
terms correspond to one new effective mass operator.  Of course there are many 
possible 13 operators.  In this work we have not performed a search over all 
possible 13 operators. Instead we study two 13 operators which are motivated 
by two complete SO(10) extensions of model 4.

Finally, we should note that we have also studied the other 8 models of ADHRS.  The
$\chi^2$ values for these models are significantly larger than those for model 4 discussed here.
 
 \section{Models 4(a,b,c) of LR}
 
In this section we analyze two models derived from complete SO(10) SUSY GUTs 
discussed recently by Lucas and S.R.\cite{lucas}.  The models were constructed 
as simple extensions of model 4 of ADHRS.  The label (a,b,c) refers to the 
different possible 22 operators which give identical Clebsch relations for the 
22 element of the Yukawa matrices.\footnote{Note, models d, e and f
have the second family $16_2$ coupled directly to $10_1$ and a heavy $16$.  If 
this coupling is as large as the third generation Yukawa coupling, then we would 
obtain excessively large flavor changing neutral current processes, such as 
$\mu \rightarrow e \gamma$.  Thus these models were not considered in \cite{lucas}.}   
However in the extension to a complete  SUSY GUT these different operators lead to 
inequivalent theories.  The  different theories are defined by the inequivalent U(1) 
quantum numbers of  the states.  When one demands ``naturalness", i.e.  includes all 
terms in the superspace potential consistent with the symmetries of the theory one  
finds an additional 13 operator for models 4a and 4c given by -- 

\begin{eqnarray} \co{13} = \\ & (a) &\;\; 16_1\  \left({\at \over
\cs{M}}\right)^3\  10_1\  
\left({\at A_2\over{\cs{M}}^2}\right)\  16_3 \nn \\ & 
(c) &\;\; 16_1\ \left({\at \over \cs{M}} \right)^3 \ 10_1 \ 
\left({A_2\over \cs{M}} \right)\  16_3 \nn 
\end{eqnarray}

 Model 4b, on the other hand, is identical to model 4 of ADHRS when states 
 with mass greater than $M_G$ are integrated out.  The results for this model
 are identical to those presented in the previous section; thus we will not 
 discuss it further. The other two models include one new effective operator 
 at the GUT scale.   The addition of this 13 operator introduces two new real 
 parameters in the Yukawa matrices at $M_G$; thus we have $n_y = 7$.  Models 
 4a and 4c differ only by the 13 operator.   The resulting Yukawa
matrices are given by --

$$Y_u=\pmatrix{ 0& C & u_u D e^{i \delta} \CR
C & 0 & -\FR1.3. B \CR
u'_u D e^{i \delta} & -\FR4.3. B & A}$$
$$Y_d=\pmatrix{ 0& -27 C & u_d D e^{i \delta} \CR
-27 C & E e^{i \phi} & \FR1.9. B \CR
u'_d D e^{i \delta} & -\FR2.9. B & A}$$
$$Y_e=\pmatrix{ 0& -27 C & u_e D e^{i \delta} \CR
-27 C & 3 E e^{i \phi} & B \CR
u'_e D e^{i \delta} & 2 B & A}$$
where the new Clebsches for models 4(a,c) are given in table \ref{t:uclebsch}.

We find that model 4a gives a best fit $\chi^2 \sim 4$ for 3 dof, while 
model 4c gives $\chi^2 \leq 1$ for 3 dof. For model 4c,  our best fit 
 with $\chi^2 = 0.168$ is found for $\mu = 160 GeV, M_{1/2} = 400 GeV, m_0 = 2900 GeV$.
 Our results are presented 
in figures [3(a - c)] for model 4a and figures [4(a - c)] for model 4c.

In order to understand why the 13 operator resolves the problems discussed in
the previous section, consider the approximate formulas (valid to 10\% and
particular to model 4c)
 for $V_{ub}/V_{cb}$,
$V_{cb}$, $\lambda_s$, $\lambda_d$, $\lambda_{\mu}$, $\lambda_e$ where the latter
are the diagonalized Yukawa eigenvalues and all are evaluated at the GUT scale.
We have \begin{eqnarray}
V_{ub}/V_{cb} & =  &{9\, C \, A \over 4\, B^2}      \\
V_{cb} & =  &  { 4\,B \over 9\, A}    \nn \\
\lambda_s & = & E  \nn  \\
\lambda_d & = &  {729 \, C^2 \over E} 
\left( 1 - {31 \over 243}{B\, D \over C \, A}
e^{-i\, \delta} \right) \nn \\
\lambda_{\mu} & = &  3\, E  \nn \\
\lambda_e & = &  {243 \, C^2 \over  E}
\left(1 - {109 \over 27}{B \, D \over C \, A} 
e^{-i \, \delta}\right)   \nn  \\
\end{eqnarray}
Recall, the contribution of the 13 operator is proportional to $D$.
Consider first the electron and down quark Yukawa couplings.  The electron
gets a significant correction from the 13 operator whereas the correction to
the down quark is much smaller.  When one adds the 13 operator to a model which
is already fit to the lepton masses, one must readjust the parameters.  In order
to keep the electron mass fixed, one needs to increase the parameter $C$ [note, in
the fits $\delta \sim 2\pi$]; the parameter $E$ is kept fixed in order not to change
the muon mass.  As a consequence the ratio $V_{ub}/V_{cb}$ increases, which allows
for smaller values of $\hat B_K$.  In addition we see that the Georgi -
Jarlskog relation is now given by
 $$ {m_s \over m_d} \approx 
{1 \over 9} \left( {1 - {109 \over 27}{B \, D \over C \, A} 
e^{-i \, \delta} \over    1 - {31 \over 243}{B\, D \over C \, A}
e^{-i\, \delta} }  \right)  {m_{\mu} \over m_e} $$
As a consequence, the ratio 1/9 is further decreased; hence Q is reduced.
Thus one new operator is able to resolve four problems.

Let us now discuss our results for model 4c. We find the preferred region 
of SUSY parameter space with a fixed value of $\,\mu \!=\! 80$GeV corresponds 
to  $\,M_{1/2} > 220$GeV and $\,m_0>300$GeV. The lower bound on $m_0$ depends 
slowly on $M_{1/2}$ while the lower bound on $M_{1/2}$ appears to be independent 
of $m_0$. These bounds are rather distinct (the $\chi^2$ value rises steeply 
when getting closer to these values) since they result from the lower 
experimental limits on sparticle masses. No observable from our list in 
\mbox{table \ref{t:observables} $\;\,$ places} 
a decisive constraint which would exclude a portion of the $(m_0,M_{1/2})$ 
parameter space at fixed $\,\mu = 80$GeV. In other words, for this value of 
$\mu$ everything in the $(m_0,M_{1/2})$ plane that has not been excluded by 
direct experimental searches, is allowed (provided the other parameters are 
subject to the optimization procedure). 

As $\mu$ increases (see figs. 4b and 4c for 
$\mu = 160,\; 240$ GeV, resp.)  the lower bound on $M_{1/2}$ 
goes down since it is determined by the LEP limits on the masses of the 
lightest chargino and neutralino. The latter become proportional to, but 
less than, $M_{1/2}$ in the limit of large $\mu$. In addition, as $\mu$ increases
the 
$\chi^2$ profile in the $m_0$ direction starts to change more smoothly, and 
the contour lines of constant $\chi^2$ shift more towards the higher values of 
$m_0$. This is explained by the fact that the constraints placed by the 
observables which receive one loop corrections enhanced by tan$\beta$ start 
to be significant. (Especially the $b$ quark mass correction and subsequently 
the correction to $M_b-M_c$.)  Since these one loop SUSY threshold corrections 
scale roughly as $ \mu \, M_{1/2}/m_0^2 \;\;  {\rm or} \;\; 
\mu \, A_{33}/ m_0^2 $ we see that with increasing $\mu$ a larger 
suppression by $m_0$ is required to keep the potentially large corrections 
under control. The effect clearly shows up in figures 4a,b,c combined. 

Our results show that the SUSY sector can make its presence visible 
in the analysis of the SM fermion mass parameters, especially those suppressed 
by tan$\beta$ at tree level, through the low energy threshold corrections. 
For $\chi^2 \sim 1$ the latter are up to 9\%. Two notes are in order at this 
point:  First, note that these corrections have the potential to be as large 
as 30-70\% \footnote{Clearly, such large corrections would not violate 
perturbativity,  since the main reason for them being large would be that 
the corresponding tree level values happen to be tan$\beta$ suppressed. That 
means that the higher order corrections would be well under control as for 
any other quantity not suppressed at tree level.}\cite{tanbetcorr}. 
It would be interesting if a window in parameter space with such large SUSY 
contributions was favored. However, our numerical optimization never disclosed 
such a region.  Second, it is interesting to note that the best fit is 
found in the region of very large $m_0$, where the effect of the SUSY threshold
corrections to fermion masses and mixings is, in fact, minimized. As a result, in
this  region, with $\chi^2 < 1$ and the SUSY corrections to fermion masses 
negligible, the effective number of degrees of freedom is
actually larger than 3, since there are 7 parameters in the Yukawa matrices
determining the 13 low energy observables in the fermion mass sector. {\em This 
means that the Yukawa sector of the selected model does actually a much better 
job than appears at first glance.}

For the opposite sign of $\mu$ there is a significant restriction from the 
process $b\rightarrow s\gamma$. As, has been noticed by many authors, the SM 
amplitude with the $W$ in the loop is sufficiently big to explain the 
measured branching ratio. In the large tan$\beta$ regime of the MSSM however, 
there are substantial partial amplitudes due to the charged Higgs and chargino 
loops. The charged Higgs amplitude is always of the same sign as the SM 
amplitude, while the sign of the chargino amplitude depends on the sign of 
$\mu A_t$. In order to get an agreement with the measured rate, the chargino 
amplitude, especially with large tan$\beta$, has to enter with the opposite 
sign, and as a consequence a specific sign of $\mu A_t$ is preferred. 
Due to the boundary conditions at the GUT scale which are used throughout 
this paper we always run into the region at the $Z$ scale where $A_t<0$, and
as a result the sign of $\mu$ is fixed.  ($\mu > 0$, in our convention.) 
With the bad sign of $\mu$, the best $\chi^2$ values 
 are about 30.\footnote{Obviously, if superpartners are {\em very} 
 heavy, the SUSY spectrum decouples and the above statement is
 incorrect. However, for large tan$\beta$, ``{\em very} heavy'' in 
 this case means deep in the TeV region which we don't consider 
 in our analysis.}  For this reason we do not quote results 
or figures for $\mu < 0$. Interestingly enough, even if we exclude the 
branching ratio for $b\rightarrow s\gamma$ from the list of observables and 
repeat the optimization procedure for $\mu < 0$, the conspiracy between the 
one loop corrections proportional to $\mu$ (see the two previous paragraphs) 
does not work very well and significantly worse 
$\chi^2$ values are obtained than in the case $\mu > 0$.

Finally note that $|\mu|$ less than about $65$GeV is excluded since 
there would inevitably be a light higgsino-like neutralino and chargino;
already ruled out by experiment. Moreover, the large values of $m_0$
selected by our fit, imply a suppression of the supersymmetric
corrections to the anomalous magnetic moment of the muon, which could
otherwise lead to very strong constraints on the allowed parameter
space in the large $\tan\beta$ region \cite{anmu}.

In tables \ref{t:model4cI} - \ref{t:model4cIII2} we give the values of the 
initial parameters which minimize $\chi^2$ for fixed particular values of $\mu, \;
m_0,\; M_{1/2}$ labelled as points I-III  marked in figures 4a-c. For each low
energy observable we quote its computed value, the partial contribution to
$\chi^2$ and the relative magnitude of the  SUSY threshold corrections [in \%].  
We also include in tables \ref{t:model4cpred} and  \ref{t:model4cpredIII} values for
the SUSY and Higgs spectra, the unitarity  triangle parameters $\rho,\,\eta$
together with the corresponding angles 
$\alpha,\; \beta$ and $\gamma$, measurable in neutral B decays, and the predicted
ratio for $m_u/m_d$. 

Note that the values for  $\alpha,\; \beta$ 
and $\gamma$ do not significantly change across different points in the 
SUSY parameter space. This fact is explicitly evident from figure 5. 
Each ``x''-symbol in this figure stands for a point anywhere in the SUSY parameter 
space with the minimum of $\chi^2$ less than 3, with 3 degrees of freedom present 
both in model 4c and model 4a. Note that, under such a restriction in model 4a, 
only the corner in the SUSY parameter space with very large $m_0$ and $M_{1/2}$ 
contributes to figure 5b. Nevertheless, these two models give similar, narrowly 
spread predictions for sin2$\alpha$ and sin2$\beta$ , grouped around the values 
0.95 and 0.52, respectively, while a much larger region is allowed for these two 
observables as a result of a general Standard Model analysis \cite{ali}.
 
 We have also checked the sensitivity of our results, for point I in fig. 4a, 
to a potential 10\% positive enhancement of $\epsilon_K$. We have found that the 
total $\chi^2$ stays the same to within a few per cent and that our predictions for 
sin2$\alpha$ (sin2$\beta$) change from 0.953 (0.513) to 0.957 (0.527).  Indeed,
 over a large range of points, we checked that the predictions for sin2$\alpha$ 
(sin2$\beta$) change by less than 1\% (3\%).  Thus our results, for model 4c, 
are not very sensitive to SUSY box corrections to $\epsilon_K$. On the other 
hand, for point II (fig. 3a) in model 4a, we find that a 10\% positive 
enhancement of $\epsilon_K$ significantly improves the agreement with 
experiment; $\chi^2$ changes from 6.99 to 4.57.  This is because in this case, 
$\hat B_K$ was a significant constraint on the model.

It is an interesting feature of most of the points in the SUSY parameter space 
that the best fits favor a rather large negative  GUT threshold 
correction $\epsilon_3$ (eqn. \ref{eq:eps3}). In figure 6a the sensitivity of 
the best fits to  different lower bounds on $\epsilon_3$ is displayed for values 
of $\mu$ and  $M_{1/2}$ fixed and $m_0$ varied. In addition table \ref{t:e3} shows 
how the selected GUT scale parameters $\alpha_G,\;M_G$ and $A$ (the 33 element 
in the Yukawa matrices) vary in the best fits at $m_0$=700GeV (point II), in 
order to compensate for changes in the lower bound on $\epsilon_3$. Our study 
shows that even small positive GUT threshold corrections 1-2\% to $\alpha_s$ 
are plausible in model 4c.  As one can see in table \ref{t:e3}, this would 
require a rather low value of the GUT scale, below $1\times 10^{16}$GeV, and 
lower $\alpha_G$. As a result, the best fit value of $\alpha_s(M_Z) \sim 0.116$ 
is getting closer to its central value 0.118, while $G_{\mu}$ is getting worse. 
On the other hand, the $A$ parameter of the Yukawa matrices is quite insensitive 
to the change in $\epsilon_3$ and to the induced changes in $\alpha_G$ and 
$M_G$, and so are the masses of the heaviest generation.

We note that our results in this regard differ from those of a recent paper by 
Pierce et al.\cite{pierce} in which a general MSSM analysis is performed. 
We believe that a significant part of the difference results from 
our assumed theoretical uncertainties in the electroweak observables.  Recall, 
in our analysis we have assumed an overall 0.5\% theoretical uncertainty ($\sigma$) 
for $M_Z$, $M_W$, and $\alpha_{EM}$ and a 1\% uncertainty for $G_{\mu}$.  As commented 
previously, we believe these uncertainties correctly account for the theoretical errors
 introduced by neglecting higher orders in perturbation theory and by the numerical 
 analysis. However for comparison, we have changed the theoretical uncertainties for 
 $M_Z$, $\alpha_{EM}$, and $G_{\mu}$ to 0.1\% and for $M_W$ to the experimental value
  of 130 MeV; the new results, for model 4c, are displayed in figure 6b and table 
  \ref{t:e3b}. In table \ref{t:e3b} we see that in this case $\alpha_s$ varies 
  significantly with changes in $\epsilon_3$.  Moreover, as $\epsilon_3$ increases 
  the quality of the fit quickly deteriorates.  This is seen graphically in figure 6b.   
  We find, for point II in model 4c with $\epsilon_3$ = -3\% (-2\%), $\chi^2$ = 7.49 (13.97) 
  with the dominant contributions to $\chi^2$ coming from $M_Z$ -- 2.11 (3.67), 
  $G_{\mu}$ -- 0.89 (1.81), $\alpha_{EM}$ -- 0.74 (1.46) and $m_b(M_b)$  -- 2.05 (3.69).   
  Thus the theoretical uncertainties for the precisely measured electroweak parameters 
  allow for quite a bit of flexibility in the values of $\epsilon_3$ and $\alpha_s(M_Z)$.

Since we have focused on fermion masses we have not discussed 
the Higgs sector  
of particular models in great detail in this work. However, in figure 7 we 
show the sensitivity of the best fits to the assumed lower bound of 
the pseudoscalar 
mass,  at fixed values of $\mu$=80GeV and $M_{1/2}$=240GeV in model 4c. 
As one can 
see, requiring a heavier pseudoscalar (A) forces the model into the 
region of larger 
$m_0$ and thus heavier squarks and sleptons.  Or in 
other words, in order to have 
light squarks and sleptons, one necessarily also has a light~A.

To understand these properties, it is instructive to analyze the
renormalization group evolution of the mass parameters. For a
top quark mass $M_t \simeq 175$ GeV, the low energy  soft
supersymmetry breaking mass parameters 
are approximately given by \cite{nonuniv},
\begin{eqnarray}
m_{H_u}^2 & \simeq & \frac{2}{3} 
\left(m_{H_u}^2(0) - m_0^2 \right) -2.5 M_{1/2}^2,
\nonumber\\
m_{H_d}^2 & \simeq & 
\frac{2}{3} \left(m_{H_d}^2(0) - m_0^2\right) -2.3 M_{1/2}^2
- \frac{0.2}{3} \left( 2 \; m_0^2 + m_{H_d}^2(0) \right),   
\nonumber\\
m_Q^2 & \simeq & \frac{5 \; m_0^2}{9} - \frac{m_{H_u}^2(0) 
+ m_{H_d}^2(0)}{9} + 5 M_{1/2}^2,
\nonumber\\
m_D^2 & \simeq & \frac{5 \;m_0^2}{9} - \frac{2 \; m_{H_d}^2(0)}{9} 
+ 5 M_{1/2}^2,
\label{eq:masses}
\end{eqnarray} 
where $m_Q$ and $m_D$ are the left and right handed sbottom mass 
parameters, and we have ignored the $A_0$ contribution, as well as
the small $m_{H_u}(m_{H_d})$ effect on
the running of $m_{H_d}(m_{H_u})$. 
Moreover, in the large $\tan\beta$ regime there is a simple relation
between the Higgs mass parameters and the pseudoscalar Higgs mass,
namely \cite{copw},
\begin{equation} 
m_A^2 \simeq m_{H_d}^2 - m_{H_u}^2 - M_Z^2.
\end{equation}
Using the fact that $m_{H_u}^2 \simeq -M_Z^2/2 - \mu^2$,
we obtain, 
\begin{equation}
m_{H_u}^2(0) - m_0^2 \simeq 4 M_{1/2}^2 - 
\frac{3}{2}\left(\mu^2 + \frac{M_Z^2}{2}\right).
\label{eq:m0m12}
\end{equation}  
Since the best fit is obtained for moderate values of the 
one-loop down quark
mass corrections, low values of the $\mu$ parameter $\mu^2 \ll m_0^2$
are preferred. 
For $M_{1/2} > \mu \simeq {\cal{O}}(M_Z)$, an approximate relation
between $m_{H_u}(0)$ and $m_0$ is obtained as a function of 
the gaugino mass parameter $M_{1/2}$. Moreover, 
\begin{equation}
0.9 \; m_{H_d}^2(0) - m_{H_u}^2(0) \simeq 0.1 \left(2 \; m_0^2  
- 3 \; M_{1/2}^2\right) 
+ \frac{3}{2}(m_A^2 + M_Z^2).
\label{eq:mA}
\end{equation}
From the above equations, the qualitative behaviour of our solutions 
may be understood. For instance, 
for $m_0 \simeq 3 M_{1/2}$, as is the case at point II 
on fig. 4a (see table V), and
$M_{1/2} > \mu,M_Z$, from Eq. (\ref{eq:m0m12}) we get that 
$m_{H_u}(0) \simeq 1.2 \; m_0$. Moreover
from Eq. (\ref{eq:mA}), we obtain that
$m_{H_d}(0)$ may vary from values of order 1.4 $m_0$, for low values 
of $m_A$, up to values of order 2 $m_0$ for larger values of $m_A$, 
without inducing a problem in the sbottom sector. Larger values of 
$m_{H_d}(0)$, however, induce lower values of the sbottom mass 
parameters, implying
unwanted large values for the gluino-induced bottom mass corrections.  
Therefore, the best fit is obtained for values of the CP-odd Higgs mass
close to its experimental bound, or, equivalently, for values of
$m_{H_d}(0)$ of the order of 1.4 $m_0$. Larger (smaller) values 
of $m_0/M_{1/2}$ imply  smaller (larger) values
of $m_{H_{u,d}}(0)/m_0$, as is clearly seen from 
Eqs. (\ref{eq:m0m12}), (\ref{eq:mA})
(see Tables III-VII). 

The relation between the CP-odd mass and $m_0$ may also be understood
from the above equations. 
For large values of the CP-odd mass, and $m_0 > M_{1/2}$, 
$m_{H_d}^2(0)$
must be significantly larger than $m_{H_u}^2(0)$. 
For these large values of $m_{H_d}^2(0)$, 
large values of $m_0^2$ are required
in order to avoid a very low value of the sbottom mass.
The results displayed in fig. 7 are just a reflection of this fact.  

To conclude this section let's discuss how the performance of models 4a-c 
changes if recent lattice results on the low value of the strange quark mass 
prove to be correct \cite{lattice:ms}. With $m_s(1GeV)=(120\pm25)MeV$ 
replacing the value $(180\pm50)MeV$ of table \ref{t:observables}, the 
$\chi^2$ value of model 4c at point II changes from 0.731/3dof to a
significantly worse value of 5.44/3dof. More than a half of it comes from 
the $m_s$ value itself, 163.4MeV at the best fit, indicating that model 
4c cannot get to such low values of the strange mass as reported by the 
lattice groups, even with the help of the fifth effective operator. The 
rest of the $\chi^2$ value is being shared by a low value for $\alpha_s(M_Z)$,
a high $V_{cb}$ and $\hat{B}_K$, and low $V_{ub}/V_{cb}$, contributing 
by less than 0.8 to the total $\chi^2$ each. We checked that this behaviour of
model 4c is typical for different points as well, and that the best  fit values of
the strange quark mass
$m_s<$ 160MeV never occurs.  Model 4a, on the contrary, favors values of
$m_s(1GeV)$ less than 180MeV. However, decreasing $m_s$ cannot cure the problems
of  this model:  high values of $\hat{B}_K$ and $V_{cb}$,
and a low value for $V_{ub}/V_{cb}$. At the same point II, these observables 
contribute  2.85, 0.75 and 1.76, respectively, to $\chi^2$ which totals 
6.82 despite values of $m_s$ as low as 136.5MeV. This represents only a marginal 
improvement to the $\chi^2$ value of 6.99 at the same point with the original
larger  experimental value of the $s$-quark mass. Finally, the performance of model
4  gets slightly worse with the low value of $m_s$: total $\chi^2$ changes 
from 14.01 to 16.16 as $m_s$ goes from 163MeV down to 156MeV in the best 
fits at point II.

\section{Discussion and Conclusions}

The results of our analysis, as well as the whole project presented here, may 
be understood from two different perspectives. 

The emphasis of this paper is to reanalyze the fermion mass sector in the context 
of the minimal SO(10) SUSY GUT in order to gain insight into the underlying 
flavor physics. From this perspective, we analyzed the best working model of 
ADHRS with four effective operators in the Yukawa sector at 
the GUT scale. The best fit allows us to assign a confidence level of about 
3-4\% to this model. Next, we found that the addition of a new operator, giving 
rise to the 13 (and 31) entries in the Yukawa matrices, may improve the performance 
of the model. Substantial improvement however, is not automatic, as has been 
evidenced with model 4a that has a confidence level $<$ 50\%  for the
best  fits in the corner of SUSY parameter space with very large $m_0$ 
and $M_{1/2}$, and large $\mu$. On the other hand, we showed that model 4c provides
an excellent fit to all 20 low energy observables, with confidence level better 
than 68\% in a large region of the allowed SUSY parameter space. We have not 
performed a complete search over all possible 13 operators. Model 4c was picked 
as a candidate model suggested by a recent formulation of a complete SO(10) SUSY
 GUT\cite{lucas}. Whether or not this particular model is close to the path 
 nature has chosen remains to be seen.  One important test will be via the CP
 violating decays of the B.  We predict a value for sin$2\alpha$ which is
 insensitive to the SUSY breaking parameters (see figure 5 and tables 
\ref{t:model4cpred} and \ref{t:model4cpredIII}),
 whereas in the SM the value of sin$2\alpha$ is unrestricted\cite{ali}.
 Another important test will come from nucleon decay rates which is
 discussed in a recent paper\cite{nucleondecay}.

As a separate matter, we considered the sensitivity of our results to different 
lower bounds for $\epsilon_3$.  For the assumed theoretical uncertainties, as given 
in table \ref{t:observables}, our analysis of model 4c indicates that the best fits 
favor a negative GUT threshold correction to $\alpha_s$ of the order -(4-5)\%.  We 
have checked for the whole $m_0$ - $M_{1/2}$ plane at fixed $\mu=80GeV$ that a 
correction of -3\% instead of -(4-5)\% can  easily be accomodated and gives basically 
the same best values of $\chi^2$ as obtained in the unrestricted analysis.  In this 
case, for $\epsilon_3 = -3$\%, the contour lines of figure 4a move almost uniformly 
towards higher values of $m_0$ by about 50 GeV. Note that the study of the 
complete SO(10) model\cite{lucas} giving rise to the effective operators of 
model 4c shows that such negative corrections to $\alpha_s(M_G)$ can easily
be obtained without any fine tuning. To understand how sensitive our results might be 
to this threshold correction we studied  a narrower region in the parameter space, 
determined by fixed $M_{1/2}$=240GeV  at the previously fixed $\mu$=80GeV (see figure 
6a and table \ref{t:e3}).   It turns out that $\epsilon_3>0$ remains a possibility, 
at the expense of lower GUT scale and $\alpha_G$,  resulting in a worse agreement with 
the low energy value of $G_{\mu}$. Fermion masses are not very sensitive to these variations.  
In addition, from table \ref{t:e3} it is clear that as $\epsilon_3$ increases, so does $\alpha_s(M_Z)$.  These results are however strongly dependent on the magnitude of our 
assumed theoretical uncertainties as discussed in the previous section and shown in 
figure 6b and table \ref{t:e3b}.

Note that model 4c is also flexible enough to 
accomodate recent analyses of the $R_b$ 
anomally and low values for $\alpha_s(M_Z)$ \cite{Rb,globfit}.
For the Higgs sector of the MSSM, the best fits 
at a number of points (e.g. also at the sample points I - III; see 
tables \ref{t:model4cpred} and \ref{t:model4cpredIII}) end up with low values 
of the pseudoscalar Higgs mass $m_A$. This preference is not strong though,
as can be inferred from figure 7 where we show how the quality of the fits 
change with the increasing lower bound on $m_A$. However, 
the low value of $m_A$, 
such as at point III2 for instance, leaves the door open for a natural 
explanation of the 1-2$\sigma$ increase in the partial 
width $Z\rightarrow b\bar{b}$, 
without ever asking for it in the course of our analysis. 
At the same time, as can be argued on general grounds, one of 
the CP even Higgs states is always almost degenerate with the 
pseudoscalar state. For $m_A<M_Z$ it is the lighter $h^0$ state, 
while for $m_A>M_Z$ it is the heavy Higgs state $H^0$.
In the latter case, the $h^0$ mass equals   $M_Z$cos$^2 2\beta\: + \:rad.\:
corrections$ where the radiative corrections are rather substantial 
in our case. 
Thus our study indicates that there are no
additional constraints on a light CP even  Higgs state between the experimental
limit of about 55GeV (for $\tan\beta \simeq 60$)
and the upper  MSSM limit of about 130GeV.

Another interesting feature of the best fits is the tendency towards lower 
values of $\alpha_s(M_Z) \sim 0.115$. As with low $m_A$, it just happens to be in 
agreement  with a potentially significant 
positive pseudoscalar contribution to 
$R_b$. The origin of rather low $\alpha_s$ 
in the best fits can be traced to the 
optimization efforts to suppress the gluino correction to an already 
large enough value for $m_b(M_Z)$. This correction has to be positive, 
since its sign is correlated with the sign of the chargino-stop partial 
amplitude to the process $b\rightarrow s\gamma$ \cite{copw}, and the latter 
is fixed in the large tan$\beta$ regime, as described in the previous section.
At the same time lower values of $\alpha_s(M_Z)$ are welcome in order to 
suppress the QCD logarithmic corrections to $b\rightarrow s\gamma$. 
Here 
the point is that with tan$\beta$ large the chargino-stop loop tends to be 
too large, most of the times outweighing by too much the $W$ and $H^+$ loops 
which contribute with the opposite sign. Thus the optimization favors lower 
$\alpha_s(M_Z)$ as the way to make up for a rather large net
$b\rightarrow s\gamma$ amplitude by minimizing the enhancement 
from the renormalization to the $M_b$ scale.

Our analysis can also be viewed from a different perspective. Let us neglect 
for a moment the underlying GUT physics and the origin of the 
Yukawa matrices at the GUT scale and view this analysis simply as an MSSM global 
fit in the large $\tan\beta$ regime. (See also \cite{globfit}.)  
>From this new 
perspective a special feature of our approach is that we run complete 
sets of fermion and sfermion mass matrices down to the $Z$ scale, instead of
just the leading 33 elements. We now try to draw some general conclusions
coming from such an analysis.  

Our results suggest that there is no narrow, 
strongly preferred region in the SUSY $(m_0,M_{1/2})$ parameter plane for 
low values of the $\mu$ parameter $\mu(M_Z)\sim 80 GeV$ .
Hence, at 
present one cannot make strong conclusions in the large tan$\beta$ 
regime about 
the masses of squarks, sleptons and gluinos, which leaves open  various 
channels for Tevatron and LEPII experiments. 

In addition, a definite statement can  be made about the sign of
$\mu$ following from the structure of the partial  amplitudes to the process
$b\rightarrow s\gamma $, as already explained in  more detail 
in the previous section. 
In the conventions we use $\mu$ has  to be positive.  Note, 
this conclusion, however,
depends on the assumption of universal squark and slepton 
masses and universal A
parameter at $M_G$ which constrains the sign of $A_t$ at $M_Z$.  

Finally, by keeping the complete 3 $\times$ 3 mass matrices 
for both fermions and 
sfermions one can study flavor dependent processes 
in a theory which fits the low 
energy data; for example, in rare $B$ and $K$ decays, 
$B - \bar B$ mixing, lepton 
flavor violating processes or even $Z \rightarrow b \bar b$.

In summary, we performed a detailed global analysis of independent SM 
parameters (plus the extra low energy data constraining the SUSY sector)
in several models based on SO(10) SUSY GUTs. At the present time, when 
direct evidence for physics beyond the Standard Model evades experimental 
observations, this kind of analysis serves as the best test of new physics 
and actually starts to compete with  SM electroweak precision tests. 
Our ultimate goal is to identify a set of effective theories, defined at a scale M 
(M = $M_{Planck},\; M_{string} \; {\rm or} \; M_{GUT}$) which accurately fit the low
energy data and thus determine the boundary conditions at the scale M for some more
fundamental theory valid above M. We have shown that this set is non-trivial; model
4c is one element of the set.  More elements of the set need to be found in order to
determine (a) whether they can be distinguished by purely low energy measurements, or (b)
whether one or more elements can be constructed as the low energy limit of a string. 

Finally,  some theoretical uncertainties in the present analysis can and should be
removed in the future.  By including the SUSY box contributions to $G_{\mu}$ we can
remove the additional 1/2\% uncertainty included in the evaluation of $\chi^2$ for this
quantity.  In addition, the SUSY box corrections to $\epsilon_K$ can be as large as
15\%\cite{goto}.  These should be included in order to remove the uncertainties in the
predictions for CP violating parameters; although these uncertainties do not appear to 
be significant at the representative points studied in model 4c.

\vskip 0.4in
\noindent {\Large\bf Acknowledgments}
\vskip 0.2in

This research was supported in part by the U.S. Department 
of Energy contract DE-ER-01545-681.  T.B. would like to thank Piotr Chankowski 
for many useful discussions and Steve Martin for clarifications.


\protect
\begin{table}[tp]
\caption{Experimental observables}
\label{t:observables}
$$
\begin{array}{|c|c|c|}
\hline
{\rm Observable} &{\rm Central \;\; value} & \sigma \\
 \hline
M_Z              &  91.186       & 0.46      \\
M_W              &  80.356       & 0.40      \\
G_{\mu}    &  1.166\cdot 10^{-5} & 1.2\cdot 10^{-7}   \\
\alpha_{EM}^{-1} &  137.04       & 0.69      \\
\alpha_s(M_Z)   &  0.118         & 0.005     \\
\rho_{new} & -0.6 \cdot 10^{-3}  & 2.6\cdot 10^{-3}   \\
\hline
M_t             &  175.0        &  6.0       \\
m_b(M_b)        &    4.26       &  0.11      \\
M_b - M_c       &    3.4        &  0.2       \\
m_s             &  180          & 50         \\
m_d/m_s         &  0.05         &  0.015     \\
Q^{-2}          &  0.00203      &  0.00020   \\
M_{\tau}        &  1.777        &  0.0089    \\
M_{\mu}         & 105.66        &  0.53      \\
M_e             &  0.5110       &  0.0026    \\
V_{us}          &  0.2205       &  0.0026    \\
V_{cb}          &  0.0392       &  0.003     \\
V_{ub}/V_{cb}   &  0.08         &  0.02      \\
\hat B_K        &  0.8          &  0.1       \\
\hline
B(b \rightarrow s \gamma) &  2.32\cdot 10^{-4} &  0.92\cdot 10^{-4}  \\
\hline
\end{array}
$$
\end{table}

\protect
\begin{table}
\caption{$u$ Clebsches for models 4(a) and 4(c)}
\label{t:uclebsch}
$$\begin{array}{|c|rrrrrr|}
\hline
\rm model & u_u & u'_u & u_d & u'_d & u_e & u'_e 
\\
\hline
a & -4/3 & 1/3 & -2 & -9 & -54 & 3 
\\
c & -4/3 & 1/3 & 2/3 & -9 & -54 & -1 
\\
\hline
\end{array}$$
\end{table}

 \protect
\begin{table}
\caption[3]{ 
{\bf Model 4 - Results at point II} \ \ (see point II on fig. 1a)\\
   \mbox{Initial parameters:   }\ \ \ \ \  
  1/$\alpha_G$ = 24.33,\ \ $M_G$ = 3.29$\cdot$10$^{16}$GeV,\ \ $\epsilon_3$ = -4.72\% , \\ 
   \makebox[5em] { }
  A = 0.734, \ \ B = 5.67$\cdot$10$^{-2}$,\ \  C = 0.865$\cdot$10$^{-4}$, \ \ 
E = 1.04$\cdot$10$^{-2}$,\ \  $\Phi$ =  1.76, \\ 
   \makebox[9em] { }
       $\mu$ = 80.00GeV,\ \  $m_0$ = 700.00GeV,\ \  $M_{1/2}$ = 240.00GeV,\\ 
   \makebox[6em] { }
$m_{H_d}/m_0$ = 1.39,\ \  $m_{H_u}/m_0$ = 1.22,\ \  $A_0$ = 387.79GeV,\ \ 
                                                    $B\mu$ =120.01GeV$^2$ 
}
\label{t:model4}
$$
\begin{array}{|c|c|c|r|}
\hline
{\rm Observable} &{\rm Computed \;\; value} & {\rm Contribution\; to\;}\chi^2 \ 
& {\rm SUSY \;corrections \;[in \%]} \\ 
\hline
M_Z              &  91.12       & \mx[5mm]{ }<0.5  &      \mx[1.8cm]{ }  \\
M_W              &  80.37       & \mx[5mm]{ }<0.5  &      \mx[1.8cm]{ }  \\
G_{\mu}   &  1.166\cdot 10^{-5} & \mx[5mm]{ }<0.5  &      \mx[1.8cm]{ }  \\
\alpha_{EM}^{-1} &  137.0       & \mx[5mm]{ }<0.5  &  1.44\mx[1.8cm]{ }  \\
\alpha_s(M_Z)    &  0.1162      & \mx[5mm]{ }<0.5  & 12.93\mx[1.8cm]{ }  \\
\rho_{new} & +1.74\cdot 10^{-4} & \mx[5mm]{ }<0.5  &      \mx[1.8cm]{ }  \\
\hline
M_t              &  173.9       & \mx[5mm]{ }<0.5  &  0.75\mx[1.8cm]{ }  \\
m_b(M_b)         &    4.360     & 0.82             &  5.14\mx[1.8cm]{ }  \\
M_b - M_c        &    3.146     & 1.61             &  7.90\mx[1.8cm]{ }  \\
m_s              &  162.6       & \mx[5mm]{ }<0.5  &  3.68\mx[1.8cm]{ }  \\
m_d/m_s          &  0.0461      & \mx[5mm]{ }<0.5  &  0.00\mx[1.8cm]{ }  \\
Q^{-2}           &  0.00173     & 2.19             &  1.66\mx[1.8cm]{ }  \\
M_{\tau}         &  1.777       & \mx[5mm]{ }<0.5  & -1.94\mx[1.8cm]{ }  \\
M_{\mu}          & 105.6        & \mx[5mm]{ }<0.5  & -1.46\mx[1.8cm]{ }  \\
M_e              &  0.5113      & \mx[5mm]{ }<0.5  & -1.46\mx[1.8cm]{ }  \\
V_{us}           &  0.2215      & \mx[5mm]{ }<0.5  &  0.00\mx[1.8cm]{ }  \\
V_{cb}           &  0.0450      & 3.79             &  1.53\mx[1.8cm]{ }  \\
V_{ub}/V_{cb}    &  0.0463      & 2.84             &  0.00\mx[1.8cm]{ }  \\
\hat B_K         &  0.9450      & 2.10             & -3.09\mx[1.8cm]{ }  \\
\hline
B(b \rightarrow s \gamma) &  2.388\cdot 10^{-4} & \mx[5mm]{ }<0.5 & \mbox{} \\
\hline
 \multicolumn{2}{|l} {\mx[2em]{ } \rm TOTAL\;\; \chi^2} 
&\multicolumn{2}{l|} {\mx[3em]{ }             14.012  }   \\
\hline
\end{array}
$$
\end{table}

 \protect
\begin{table}
\caption[4]{ 
{\bf Model 4c - Results at point I} \ \ (see point I on fig. 4a)\\
   \mbox{Initial parameters:   }\ \ \ \ \  
  1/$\alpha_G$ = 24.43,\ \ $M_G$ = 2.50$\cdot$10$^{16}$GeV,\ \ $\epsilon_3$ = -4.76\% , \\ 
  A = 0.764, B = 5.26$\cdot$10$^{-2}$, C = 1.10$\cdot$10$^{-4}$, 
D = 4.63$\cdot$10$^{-4}$, $\delta$ = 5.70, 
E = 1.25$\cdot$10$^{-2}$, $\Phi$ =  1.07, \\ 
   \makebox[9em] { }
       $\mu$ = 80.00GeV,\ \  $m_0$ = 400.00GeV,\ \  $M_{1/2}$ = 280.00GeV,\\ 
   \makebox[6em] { }
$m_{H_d}/m_0$ = 1.77,\ \  $m_{H_u}/m_0$ = 1.59,\ \  $A_0$ = 322.21GeV,\ \ 
                                                    $B\mu$ = 120.00GeV$^2$ 
}
\label{t:model4cI}
$$
\begin{array}{|c|c|c|r|}
\hline
{\rm Observable} &{\rm Computed \;\; value} & {\rm Contribution\; to\;}\chi^2 \ 
& {\rm SUSY \;corrections \;[in \%]} \\ 
\hline
M_Z              &  91.12       & \mx[5mm]{ }<0.1  &      \mx[1.8cm]{ }  \\
M_W              &  80.34       & \mx[5mm]{ }<0.1  &      \mx[1.8cm]{ }  \\
G_{\mu}   &  1.164\cdot 10^{-5} & \mx[5mm]{ }<0.1  &      \mx[1.8cm]{ }  \\
\alpha_{EM}^{-1} &  136.9       & \mx[5mm]{ }<0.1  &  1.25\mx[1.8cm]{ }  \\
\alpha_s(M_Z)    &  0.1132      & 0.93             & 12.66\mx[1.8cm]{ }  \\
\rho_{new} &+9.87\cdot 10^{-5}  & \mx[5mm]{ }<0.1  &      \mx[1.8cm]{ }  \\
\hline
M_t              &  173.5       & \mx[5mm]{ }<0.1  &  0.93\mx[1.8cm]{ }  \\
m_b(M_b)         &    4.311     & 0.22             &  4.96\mx[1.8cm]{ }  \\
M_b - M_c        &    3.499     & 0.24             &  6.64\mx[1.8cm]{ }  \\
m_s              &  184.6       & \mx[5mm]{ }<0.1  &  4.95\mx[1.8cm]{ }  \\
m_d/m_s          &  0.0496      & \mx[5mm]{ }<0.1  &  0.00\mx[1.8cm]{ }  \\
Q^{-2}           &  0.00205     & \mx[5mm]{ }<0.1  &  2.01\mx[1.8cm]{ }  \\
M_{\tau}         &  1.776       & \mx[5mm]{ }<0.1  & -4.00\mx[1.8cm]{ }  \\
M_{\mu}          & 105.7        & \mx[5mm]{ }<0.1  & -2.76\mx[1.8cm]{ }  \\
M_e              &  0.5110      & \mx[5mm]{ }<0.1  & -2.76\mx[1.8cm]{ }  \\
V_{us}           &  0.2205      & \mx[5mm]{ }<0.1  &  0.00\mx[1.8cm]{ }  \\
V_{cb}           &  0.0406      & 0.21             &  2.49\mx[1.8cm]{ }  \\
V_{ub}/V_{cb}    &  0.0744      & \mx[5mm]{ }<0.1  &  0.00\mx[1.8cm]{ }  \\
\hat B_K         &  0.8196      & \mx[5mm]{ }<0.1  & -5.05\mx[1.8cm]{ }  \\
\hline
B(b \rightarrow s \gamma) &  2.575\cdot 10^{-4} & \mx[5mm]{ }<0.1 & \mbox{} \\
\hline
 \multicolumn{2}{|l} {\mx[2.0em]{ } \rm TOTAL\;\; \chi^2} 
&\multicolumn{2}{l|} {\mx[3.5em]{ }              2.039  }   \\
\hline
\end{array}
$$
\end{table}

\protect
\begin{table}
\caption[5]{ 
{\bf Model 4c - Results at point II} \ \ (see point II on fig. 4a)\\
   \mbox{Initial parameters:   }\ \ \ \ \  
  1/$\alpha_G$ = 24.36,\ \ $M_G$ = 3.17$\cdot$10$^{16}$GeV,\ \ $\epsilon_3$ = -4.89\% , \\ 
  A = 0.807, B = 5.44$\cdot$10$^{-2}$, C = 1.15$\cdot$10$^{-4}$, 
D = 4.94$\cdot$10$^{-4}$, $\delta$ = 5.71, 
E = 1.31$\cdot$10$^{-2}$, $\Phi$ =  1.04, \\ 
   \makebox[9em] { }
       $\mu$ = 80.00GeV,\ \  $m_0$ = 700.00GeV,\ \  $M_{1/2}$ = 240.00GeV,\\ 
   \makebox[6em] { }
$m_{H_d}/m_0$ = 1.42,\ \  $m_{H_u}/m_0$ = 1.24,\ \  $A_0$ = 458.35GeV,\ \ 
                                                    $B\mu$ = 120.66GeV$^2$ 
}
\label{t:model4cII}
$$
\begin{array}{|c|c|c|r|}
\hline
{\rm Observable} &{\rm Computed \;\; value} & {\rm Contribution\; to\;}\chi^2 \ 
& {\rm SUSY \;corrections \;[in \%]} \\ 
\hline
M_Z              &  91.12       & 0.02             &      \mx[1.8cm]{ }  \\
M_W              &  80.38       & \mx[5mm]{ }<0.02 &      \mx[1.8cm]{ }  \\
G_{\mu}   &  1.166\cdot 10^{-5} & \mx[5mm]{ }<0.02 &      \mx[1.8cm]{ }  \\
\alpha_{EM}^{-1} &  137.0       & \mx[5mm]{ }<0.02 &  1.43\mx[1.8cm]{ }  \\
\alpha_s(M_Z)    &  0.1151      & 0.34             & 12.78\mx[1.8cm]{ }  \\
\rho_{new} &+1.87\cdot 10^{-4}  & 0.09             &      \mx[1.8cm]{ }  \\
\hline
M_t              &  175.7       & \mx[5mm]{ }<0.02 &  0.74\mx[1.8cm]{ }  \\
m_b(M_b)         &    4.287     & 0.06             &  5.43\mx[1.8cm]{ }  \\
M_b - M_c        &    3.440     & 0.04             &  7.56\mx[1.8cm]{ }  \\
m_s              &  189.0       & 0.03             &  3.68\mx[1.8cm]{ }  \\
m_d/m_s          &  0.0502      & \mx[5mm]{ }<0.02 &  0.00\mx[1.8cm]{ }  \\
Q^{-2}           &  0.00204     & \mx[5mm]{ }<0.02 &  1.78\mx[1.8cm]{ }  \\
M_{\tau}         &  1.776       & \mx[5mm]{ }<0.02 & -2.08\mx[1.8cm]{ }  \\
M_{\mu}          & 105.7        & \mx[5mm]{ }<0.02 & -1.50\mx[1.8cm]{ }  \\
M_e              &  0.5110      & \mx[5mm]{ }<0.02 & -1.50\mx[1.8cm]{ }  \\
V_{us}           &  0.2205      & \mx[5mm]{ }<0.02 &  0.00\mx[1.8cm]{ }  \\
V_{cb}           &  0.0400      & 0.07             &  1.58\mx[1.8cm]{ }  \\
V_{ub}/V_{cb}    &  0.0772      & \mx[5mm]{ }<0.02 &  0.00\mx[1.8cm]{ }  \\
\hat B_K         &  0.8140      & \mx[5mm]{ }<0.02 & -3.18\mx[1.8cm]{ }  \\
\hline
B(b \rightarrow s \gamma) &  2.382\cdot 10^{-4} & \mx[5mm]{ }<0.02 & \mbox{} \\
\hline
 \multicolumn{2}{|l} {\mx[2.0em]{ } \rm TOTAL\;\; \chi^2} 
&\multicolumn{2}{l|} {\mx[3.5em]{ }             0.7306}   \\
\hline
\end{array}
$$
\end{table}

\protect
\begin{table}
\caption[6]{ 
{\bf Model 4c - Results at point III1} \ \ (see point III1 on fig. 4b)\\
   \mbox{Initial parameters:   }\ \ \ \ \  
  1/$\alpha_G$ = 24.51,\ \ $M_G$ = 3.33$\cdot$10$^{16}$GeV,\ \ $\epsilon_3$ = -4.34\% , \\ 
  A = 0.852, B = 5.63$\cdot$10$^{-2}$, C = 1.21$\cdot$10$^{-4}$, 
D = 5.06$\cdot$10$^{-4}$, $\delta$ = 5.70, 
E = 1.36$\cdot$10$^{-2}$, $\Phi$ =  1.02, \\ 
   \makebox[9em] { }
       $\mu$ = 160.00GeV,\ \  $m_0$ = 1400.00GeV,\ \  $M_{1/2}$ = 170.00GeV,\\ 
   \makebox[6em] { }
$m_{H_d}/m_0$ = 1.33,\ \  $m_{H_u}/m_0$ = 1.14,\ \  $A_0$ = -982.43GeV,\ \ 
                                                    $B\mu$ = 123.72GeV$^2$ 
}
\label{t:model4cIII1}
$$
\begin{array}{|c|c|c|r|}
\hline
{\rm Observable} &{\rm Computed \;\; value} & {\rm Contribution\; to\;}\chi^2 \ 
& {\rm SUSY \;corrections \;[in \%]} \\ 
\hline
M_Z              &  91.12       & 0.02             &      \mx[1.8cm]{ }  \\
M_W              &  80.38       & \mx[5mm]{ }<0.02 &      \mx[1.8cm]{ }  \\
G_{\mu}   &  1.166\cdot 10^{-5} & \mx[5mm]{ }<0.02 &      \mx[1.8cm]{ }  \\
\alpha_{EM}^{-1} &  137.0       & \mx[5mm]{ }<0.02 &  1.79\mx[1.8cm]{ }  \\
\alpha_s(M_Z)    &  0.1154      & 0.26             & 12.89\mx[1.8cm]{ }  \\
\rho_{new} &+1.50\cdot 10^{-4}  & 0.08             &      \mx[1.8cm]{ }  \\
\hline
M_t              &  176.9       & 0.10             &  0.62\mx[1.8cm]{ }  \\
m_b(M_b)         &    4.275     & \mx[5mm]{ }<0.02 &  6.27\mx[1.8cm]{ }  \\
M_b - M_c        &    3.429     & 0.02             &  8.76\mx[1.8cm]{ }  \\
m_s              &  188.3       & 0.03             &  2.85\mx[1.8cm]{ }  \\
m_d/m_s          &  0.0512      & \mx[5mm]{ }<0.02 &  0.00\mx[1.8cm]{ }  \\
Q^{-2}           &  0.00203     & \mx[5mm]{ }<0.02 &  1.68\mx[1.8cm]{ }  \\
M_{\tau}         &  1.777       & \mx[5mm]{ }<0.02 & -1.41\mx[1.8cm]{ }  \\
M_{\mu}          & 105.7        & \mx[5mm]{ }<0.02 & -0.82\mx[1.8cm]{ }  \\
M_e              &  0.5110      & \mx[5mm]{ }<0.02 & -0.81\mx[1.8cm]{ }  \\
V_{us}           &  0.2205      & \mx[5mm]{ }<0.02 &  0.01\mx[1.8cm]{ }  \\
V_{cb}           &  0.0397      & 0.03             &  1.62\mx[1.8cm]{ }  \\
V_{ub}/V_{cb}    &  0.0797      & \mx[5mm]{ }<0.02 &  0.00\mx[1.8cm]{ }  \\
\hat B_K         &  0.8000      & \mx[5mm]{ }<0.02 & -3.26\mx[1.8cm]{ }  \\
\hline
B(b \rightarrow s \gamma) &  2.311\cdot 10^{-4} & \mx[5mm]{ }<0.02 & \mbox{} \\
\hline
 \multicolumn{2}{|l} {\mx[2.0em]{ } \rm TOTAL\;\; \chi^2} 
&\multicolumn{2}{l|} {\mx[3.5em]{ }             0.5868}   \\
\hline
\end{array}
$$
\end{table}

\protect
\begin{table}
\caption[7]{ 
{\bf Model 4c - Results at point III2} \ \ (see point III2 on fig. 4c)\\
   \mbox{Initial parameters:   }\ \ \ \ \  
  1/$\alpha_G$ = 24.65,\ \ $M_G$ = 2.88$\cdot$10$^{16}$GeV,\ \ $\epsilon_3$ = -4.45\% , \\ 
  A = 0.888, B = 5.89$\cdot$10$^{-2}$, C = 1.23$\cdot$10$^{-4}$, 
D = 5.69$\cdot$10$^{-4}$, $\delta$ = 5.74, 
E = 1.40$\cdot$10$^{-2}$, $\Phi$ =  1.02, \\ 
   \makebox[9em] { }
       $\mu$ = 240.00GeV,\ \  $m_0$ = 1400.00GeV,\ \  $M_{1/2}$ = 170.00GeV,\\ 
   \makebox[6em] { }
$m_{H_d}/m_0$ = 1.33,\ \  $m_{H_u}/m_0$ = 1.14,\ \  $A_0$ = -1079.39GeV,\ \ 
                                                    $B\mu$ = 128.50GeV$^2$ 
}
\label{t:model4cIII2}
$$
\begin{array}{|c|c|c|r|}
\hline
{\rm Observable} &{\rm Computed \;\; value} & {\rm Contribution\; to\;}\chi^2 \ 
& {\rm SUSY \;corrections \;[in \%]} \\ 
\hline
M_Z              &  91.12       & \mx[5mm]{ }<0.10 &      \mx[1.8cm]{ }  \\
M_W              &  80.35       & \mx[5mm]{ }<0.10 &      \mx[1.8cm]{ }  \\
G_{\mu}   &  1.165\cdot 10^{-5} & \mx[5mm]{ }<0.10 &      \mx[1.8cm]{ }  \\
\alpha_{EM}^{-1} &  136.9       & \mx[5mm]{ }<0.10 &  1.85\mx[1.8cm]{ }  \\
\alpha_s(M_Z)    &  0.1124      & 1.25             & 12.53\mx[1.8cm]{ }  \\
\rho_{new} &+4.60\cdot 10^{-5}  & \mx[5mm]{ }<0.10 &      \mx[1.8cm]{ }  \\
\hline
M_t              &  176.5       & \mx[5mm]{ }<0.10 &  0.61\mx[1.8cm]{ }  \\
m_b(M_b)         &    4.306     & 0.18             &  9.73\mx[1.8cm]{ }  \\
M_b - M_c        &    3.486     & 0.19             & 13.45\mx[1.8cm]{ }  \\
m_s              &  178.5       & \mx[5mm]{ }<0.10 &  4.09\mx[1.8cm]{ }  \\
m_d/m_s          &  0.0500      & \mx[5mm]{ }<0.10 &  0.00\mx[1.8cm]{ }  \\
Q^{-2}           &  0.00203     & \mx[5mm]{ }<0.10 &  1.91\mx[1.8cm]{ }  \\
M_{\tau}         &  1.776       & \mx[5mm]{ }<0.10 & -1.93\mx[1.8cm]{ }  \\
M_{\mu}          & 105.7        & \mx[5mm]{ }<0.10 & -1.06\mx[1.8cm]{ }  \\
M_e              &  0.5110      & \mx[5mm]{ }<0.10 & -1.06\mx[1.8cm]{ }  \\
V_{us}           &  0.2205      & \mx[5mm]{ }<0.10 &  0.01\mx[1.8cm]{ }  \\
V_{cb}           &  0.0402      & 0.11             &  1.73\mx[1.8cm]{ }  \\
V_{ub}/V_{cb}    &  0.0777      & \mx[5mm]{ }<0.10 &  0.00\mx[1.8cm]{ }  \\
\hat B_K         &  0.7949      & \mx[5mm]{ }<0.10 & -3.49\mx[1.8cm]{ }  \\
\hline
B(b \rightarrow s \gamma) &  2.269\cdot 10^{-4} & \mx[5mm]{ }<0.10 & \mbox{} \\
\hline
 \multicolumn{2}{|l} {\mx[2.0em]{ } \rm TOTAL\;\; \chi^2} 
&\multicolumn{2}{l|} {\mx[3.5em]{ }             1.9477}   \\
\hline
\end{array}
$$
\end{table}

\protect
\begin{table}
\caption[8]{ 
{\bf Model 4c - More Results at points I and II} \ \ 
                              (see points I and II on fig. 4a)\\
In Higgs sector, we quote all masses with one loop corrections included. 
For superpartners, masses are at tree level. \\
For squark and slepton masses, the first two columns are for the third family 
which are significantly split, while the third and fourth columns are mean 
values for the nearly degenerate states of the second and first families, respectively.}

\label{t:model4cpred}
$$\begin{array}{|l|crrr|crrr|}
\hline
{\rm Observable} &\multicolumn{4}{|c|}{\rm Predictions\; at\; point\; I} 
                 &\multicolumn{4}{|c|}{\rm Predictions\; at\; point\; II}  \\
\hline
m_u/m_d\,|_{1GeV} &   0.409 &  &  &            &  0.440     &  & &     \\
\hline
sin^2\theta_W^{(MSSM)}   & 0.2342    & & &     &  0.2334    &  & &     \\
sin^2\theta_W^{(SM)}     & 0.2323    & & &     &  0.2320    &  & &     \\
\hline
\sin 2\alpha  & 0.953       &  &  &            &  0.966     &  & &     \\
\sin 2\beta   & 0.513       &  &  &            &  0.522     &  & &     \\
\sin \gamma   & 0.935       &  &  &            &  0.928     &  & &     \\
\rho          &-0.120       &  &  &            & -0.130     &  & &     \\
\eta          & 0.316       &  &  &            &  0.325     &  & &     \\
\hline
{\rm tan}\beta           & 52.77&       & &     & 54.38&       & & \\
CP\;{\rm even\;Higgses}  & 73.43& 110.81& &     & 75.24& 110.35& & \\
CP\;{\rm odd\;Higgs}     & 73.45&       & &     & 75.25&       & & \\
{\rm charged\;Higgs}     &115.65&       & &     &116.92&       & & \\ 
\hline
{\rm gluino}     & 725              &  & &        & 630          &  & &     \\
{\rm charginos}  & 70  & 260 &     &              &  67 & 232 &     &       \\
{\rm neutralinos}& 52  &  95 & 129\mx[1mm]{} & 260\mx[1mm]{}
                 & 48  &  97 & 113\mx[1mm]{} & 232\mx[1mm]{}   \\
\hline
{\rm up\;squarks}       &  474 & 614 & \mx[3mm]{ }747 &  771  
                        &  487 & 603 & \mx[3mm]{ }887 &  905 \\
{\rm down\;squarks}     &  510 & 552 & \mx[3mm]{ }749 &  775 
                        &  513 & 550 & \mx[3mm]{ }893 &  909 \\
{\rm charged\;sleptons} &   66 & 338 & \mx[3mm]{ }423 &  448
                        &  294 & 550 & \mx[3mm]{ }716 &  718 \\
{\rm sneutrinos}        &  327 &     & \mx[3mm]{ }440 &  440  
                        &  544 &     & \mx[3mm]{ }713 &  714 \\
\hline
\end{array}$$
\end{table}

\protect
\begin{table}
\caption[9]{ 
{\bf Model 4c - More Results at points III1 and III2} \ \ 
                         (see points III1 and III2 on figures 4b and 4c )\\
Higgs, squark and slepton masses are treated in the same way as 
for the results quoted on Table VIII.}
\label{t:model4cpredIII}
$$\begin{array}{|l|crrr|crrr|}
\hline
{\rm Observable} &\multicolumn{4}{|c|}{\rm Predictions\; at\; point\; III1} 
                 &\multicolumn{4}{|c|}{\rm Predictions\; at\; point\; III2}  \\
\hline
m_u/m_d\,|_{1GeV} &   0.475 &  &  &            &  0.432     &  & &     \\
\hline
sin^2\theta_W^{(MSSM)}   & 0.2337    & & &     &  0.2344    &  & &     \\
sin^2\theta_W^{(SM)}     & 0.2325    & & &     &  0.2331    &  & &     \\
\hline
\sin 2\alpha  & 0.973       &  &  &            &  0.977     &  & &     \\
\sin 2\beta   & 0.531       &  &  &            &  0.516     &  & &     \\
\sin \gamma   & 0.925       &  &  &            &  0.918     &  & &     \\
\rho          &-0.138       &  &  &            & -0.140     &  & &     \\
\eta          & 0.334       &  &  &            &  0.323     &  & &     \\
\hline
{\rm tan}\beta           & 55.39&       & &     & 55.86&       & & \\
CP\;{\rm even\;Higgses}  & 71.70& 112.36& &     & 66.70& 112.05& & \\
CP\;{\rm odd\;Higgs}     & 71.70&       & &     & 66.70&       & & \\
{\rm charged\;Higgs}     &115.43&       & &     &112.22&       & & \\ 
\hline
{\rm gluino}     & 447              &  & &        & 440          &  & &     \\
{\rm charginos}  & 101 & 218 &     &              & 122 & 274 &     &       \\
{\rm neutralinos}&  62 & 105 & 174\mx[1mm]{} & 217\mx[1mm]{}
                 &  67 & 122 & 251\mx[1mm]{} & 272\mx[1mm]{}   \\
\hline
{\rm up\;squarks}       &  558 & 677 & \mx[3mm]{}1443 & 1460  
                        &  546 & 666 & \mx[3mm]{}1440 & 1458 \\
{\rm down\;squarks}     &  597 & 641 & \mx[3mm]{}1459 & 1462 
                        &  574 & 639 & \mx[3mm]{}1457 & 1460 \\
{\rm charged\;sleptons} &  600 &1062 & \mx[3mm]{}1397 & 1418
                        &  564 &1051 & \mx[3mm]{}1397 & 1418 \\
{\rm sneutrinos}        & 1052 &     & \mx[3mm]{}1394 & 1395  
                        & 1047 &     & \mx[3mm]{}1394 & 1395 \\
\hline
\end{array}$$
\end{table}

\protect
\begin{table}[tp]
\caption [10] {
{\bf Model 4c - Varying the Lower Bound on $\epsilon_3$ at point II} \ \ 
Best fit values of selected GUT parameters and low energy observables are 
displayed for different lower bounds on $\epsilon_3$. Note that the 
optimization procedure pushes $\epsilon_3$ to its lower limit. The first 
line corresponds to the unrestricted case and the corresponding values 
are taken over from table \ref{t:model4cII} for the sake of completeness. }

\label{t:e3}
$$
\begin{array}{|c|c|c|c||c|c|c|c||c|}
\hline
\epsilon_3 &\;\;\;1/\alpha_G\;\;\; &M_G\times 10^{-16} & A &\alpha_s(M_Z) &
G_{\mu}\times 10^5 &M_{top} &m_b(M_b) & \chi^2_{total} \\
\hline
-4.89\%    & 24.362 & 3.17& 0.807 & 0.1151 & 1.166 & 175.7 & 4.287&  0.731 \\
-3.00\%    & 24.628 & 2.10& 0.808 & 0.1154 & 1.161 & 176.1 & 4.291&  0.959 \\
-2.00\%    & 24.767 & 1.69& 0.802 & 0.1156 & 1.159 & 176.1 & 4.295&  1.279 \\
-1.00\%    & 24.907 & 1.37& 0.813 & 0.1158 & 1.157 & 176.6 & 4.295&  1.694 \\
 0.00\%    & 25.044 & 1.11& 0.818 & 0.1160 & 1.155 & 177.0 & 4.296&  2.244 \\
+1.00\%    & 25.177 & 0.89& 0.794 & 0.1162 & 1.153 & 176.4 & 4.304&  2.944 \\
+2.00\%    & 25.314 & 0.73& 0.810 & 0.1164 & 1.151 & 177.1 & 4.302&  3.675 \\
+3.00\%    & 25.450 & 0.59& 0.827 & 0.1166 & 1.149 & 177.9 & 4.303&  4.583 \\
+4.00\%    & 25.578 & 0.48& 0.797 & 0.1167 & 1.147 & 177.1 & 4.309&  5.534 \\
\hline
\end{array}
$$
\end{table}

\protect
\begin{table}[bp]
\caption [11] {

{\bf Model 4c -  Sensitivity of the results of table \ref{t:e3} to the 
magnitude of the theoretical uncertainties} \ \ 
In comparison to table \ref{t:e3}, the best fit values of the same 
GUT parameters and low energy observables are 
displayed for different lower bounds on $\epsilon_3$, but this time 
for the uncertainty $\sigma$=0.1\% for $M_Z,\: G_{\mu}$ and 
$\alpha_{em}$, and the experimental value $\sigma$=130MeV for $M_W$.
($\sigma$=0.5\% for $M_Z,\: M_W$ and $\alpha_{em}$,
and $\sigma$=1.0\% for $G_{\mu}$ was assumed in table \ref{t:e3}, as 
well as throughout this paper.) The first line corresponds to the case 
when $\epsilon_3$ is unrestricted.}

\label{t:e3b}
$$
\begin{array}{|c|c|c|c||c|c|c|c||c|}
\hline
\epsilon_3 &\;\;\;1/\alpha_G\;\;\; &M_G\times 10^{-16} & A &\alpha_s(M_Z) &
G_{\mu}\times 10^5 &M_{top} &m_b(M_b) & \chi^2_{total} \\
\hline
-5.36\%    & 24.295 & 3.53& 0.807 & 0.1150 & 1.166 & 175.6 & 4.286&  1.20 \\
-3.00\%    & 24.494 & 2.55& 0.808 & 0.1186 & 1.165 & 177.4 & 4.418&  7.49 \\
-2.00\%    & 24.579 & 2.21& 0.803 & 0.1201 & 1.165 & 177.9 & 4.471& 13.97 \\
 0.00\%    & 24.759 & 1.64& 0.818 & 0.1229 & 1.164 & 179.6 & 4.560& 34.02 \\
\hline
\end{array}
$$
\end{table}

\newpage
\noindent {\bf FIGURE CAPTIONS}
\vskip 0.5cm

\noindent {\bf Figure 1.}

\noindent 
Model 4 global analysis results in the $m_0\,-\,M_{1/2}$ plane,
for a fixed value of the Higgs parameter \\
a) $\mu(M_Z)$ = 80 GeV , \\
b) $\mu(M_Z)$ =160 GeV , \\
c) $\mu(M_Z)$ =240 GeV . \\
Solid (dashed, double-dash-dotted) lines represent contour lines 
of constant \mbox{$\,\chi^2\,=\,$ 15 (14,13) / 5dof.} \\
The star on figure 1a marks the point ($m_0$=700GeV, $M_{1/2}$=240GeV). 
In the text it is referred to as point II and the corresponding 
results obtained at this point are listed in table \ref{t:model4}.

\vskip 0.3cm

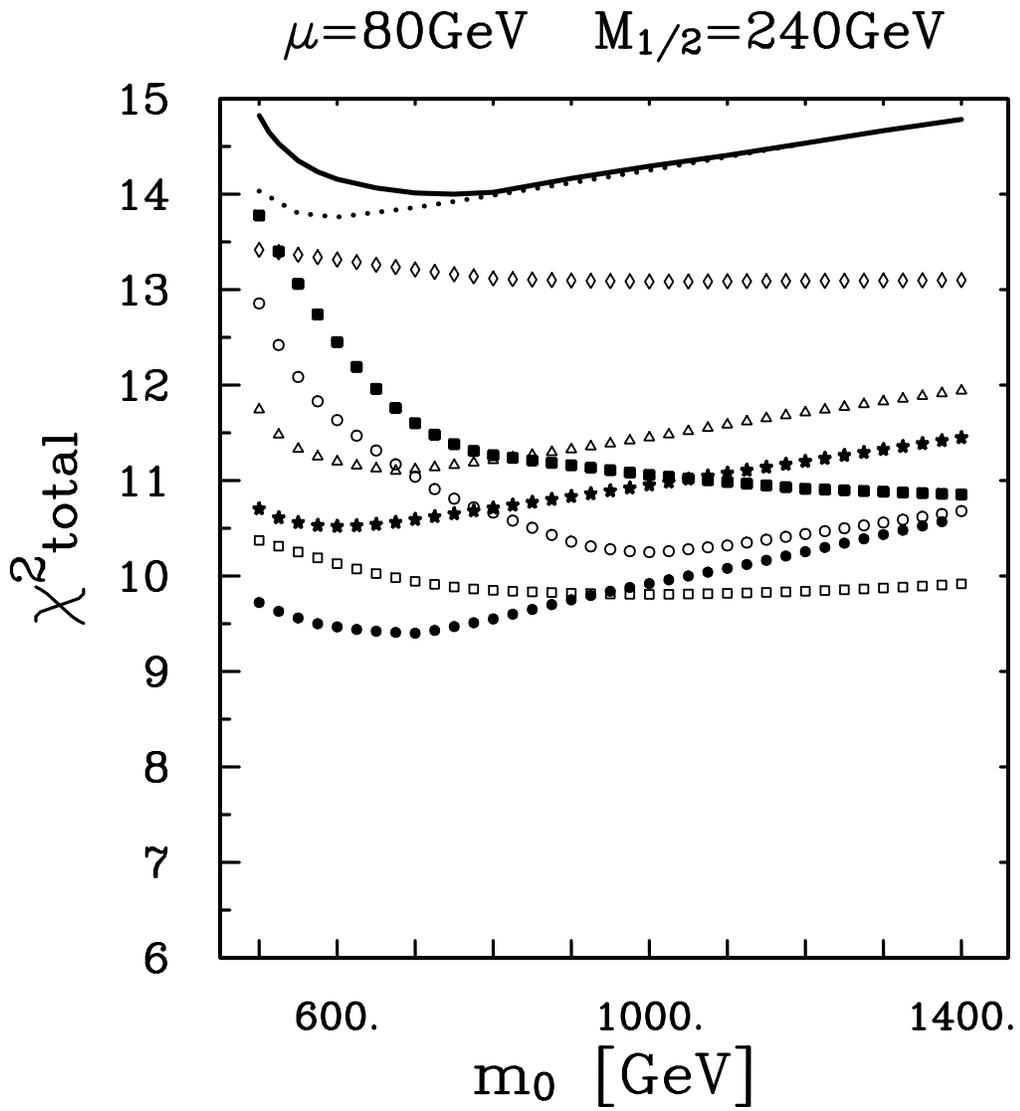
\noindent {\bf Figure 2.}

\noindent 
Results for Model 4 with one observable removed from 
the $\chi^2$ function (the corresponding standard deviation of 
that observable is inflated by a large factor), at fixed values 
of the Higgs parameter $\mu(M_Z)$=80GeV and $M_{1/2}$=240GeV.
Eight observables tried, (and their values obtained at the minimum of  
$\chi^2$) are as follows: 
$\alpha_s$                 (0.1144)                - dots, 
$BR(b\rightarrow s\gamma)$ (2.29$\cdot$ 10$^{-2}$) - diamonds, 
$M_b-M_c$                  (2.84GeV)               - solid squares, 
$1/Q^2$                    (1.50$\cdot$ 10$^{-3}$) - open circles, 
$V_{ub}/V_{cb}$            (0.045)                 - triangles, 
$m_b(M_b)$                 (4.68GeV)               - stars, 
$B_K$                      (1.12)                  - open squares, 
$V_{cb}$                   (0.0465)                - solid circles.
For comparison, the thick solid line represents the best $\chi^2$ 
values of Model 4 with all observables present with their 
regular standard deviations as quoted in table \ref{t:observables}\@. 
No significant improvement of Model 4 is observed. Recall that  
leaving out an observable from the $\chi^2$ function means 
reducing the number of degrees of freedom by one.

\vskip 0.3cm

\noindent {\bf Figure 3.}

\noindent 
Model 4a global analysis results in the $m_0\,-\,M_{1/2}$ plane,
for a fixed value of the Higgs parameter \\
a) $\mu(M_Z)$ = 80 GeV , \\
b) $\mu(M_Z)$ =160 GeV , \\
c) $\mu(M_Z)$ =240 GeV . \\
Solid (dashed, double-dash-dotted) lines represent contour lines 
of constant \mbox{$\,\chi^2\,=\,$ 6 (4,3) / 3dof.} \\

\vskip 0.3cm

\noindent {\bf Figure 4.}

\noindent 
Model 4c global analysis results in the $m_0\,-\,M_{1/2}$ plane,
for a fixed value of the Higgs parameter \\
a) $\mu(M_Z)$ = 80 GeV , \\
b) $\mu(M_Z)$ =160 GeV , \\
c) $\mu(M_Z)$ =240 GeV . \\
Solid (double-dash-dotted, dotted) lines represent contour lines 
of constant \mbox{$\,\chi^2\,=\,$ 6 (3,1) / 3dof}. 
The stars mark the points (400,280), (700,240) and (1400,170). 
In the text, these are referred to as points I, II, III1 (point III at 
$\mu$=160GeV) and III2 (point III at $\mu$=240GeV). 
The corresponding results  and predictions obtained at these 
points are listed in tables \ref{t:model4cI} through \ref{t:model4cpredIII}.

\vskip 0.3cm

\noindent {\bf Figure 5.}

\noindent 
Stability of the predictions for sin2$\alpha$ and  
sin2$\beta$, the parameters of the unitarity triangle. ``x'' symbols 
denote the points predicted by \\
a) model 4c ,\\
b) model 4a  \\
at those regions in the SUSY parameter space where $\chi^2<\;$3 / 3dof . 
Dots in the figures represent the boundary of the allowed region resulting 
from a combined general Standard Model fit \cite{ali} for $f_{B_d} = 200 \pm
 40 MeV  ,\;\; B_{B_d} =  1.0 
,\;\; \hat B_K = 0.75 \pm 0.10 $.

\vskip 0.3cm

\noindent {\bf Figure 6a.}

\noindent 
The dependence of the quality of the fit on the GUT threshold correction 
$\epsilon_3$, in model 4c. The thick solid line represents the results 
of model 4c with $\epsilon_3$ unrestricted. The dotted line constrains 
$\epsilon_3$ to be positive (and the optimization procedure always settles 
$\epsilon_3$ very close to zero). Curves with solid squares, open circles 
and stars correspond to the results of the optimization performed with 
negative lower limit on $\epsilon_3$ -3\%, -2\% and -1\%, respectively. 
Curves with open triangles, solid circles, open squares and solid triangles 
correspond to the cases with lower bound on $\epsilon_3$ greater than zero, 
namely 1\%, 2\%, 3\% and 4\%, respectively. In all cases the optimization 
procedure tends to yield the values of $\epsilon_3$ very close to the lower bound.
Table \ref{t:e3} displays the effect of different $\epsilon_3$ lower bounds 
on various GUT parameters and low energy observables at point II 
(i.e. at $m_0$=700GeV ).


\vskip 0.3cm

\noindent {\bf Figure 6b.}

\noindent 
The same as in figure 6a, with a tighter theoretical uncertainty on 
precisely measured electroweak observables. In this figure the uncertainty 
for $M_Z,\: G_{\mu}$ and $\alpha_{em}$ is set to 0.1\% and the actual 
experimental error 130MeV is assigned to $\sigma(M_W)$; as opposed to 
the uncertainty of 0.5\% for $M_Z,\: M_W$ and $\alpha_{em}$, and 
$\sigma(G_{\mu})$=1.0\% , 
assumed in figure 6a. As before, in all restricted cases the optimization 
procedure tends to yield the values of $\epsilon_3$ very close to the lower 
bound. Table \ref{t:e3b} displays the effect of different $\epsilon_3$ lower 
bounds on various GUT parameters and low energy observables at point II 
(i.e. at $m_0$=700GeV ).

\vskip 0.3cm

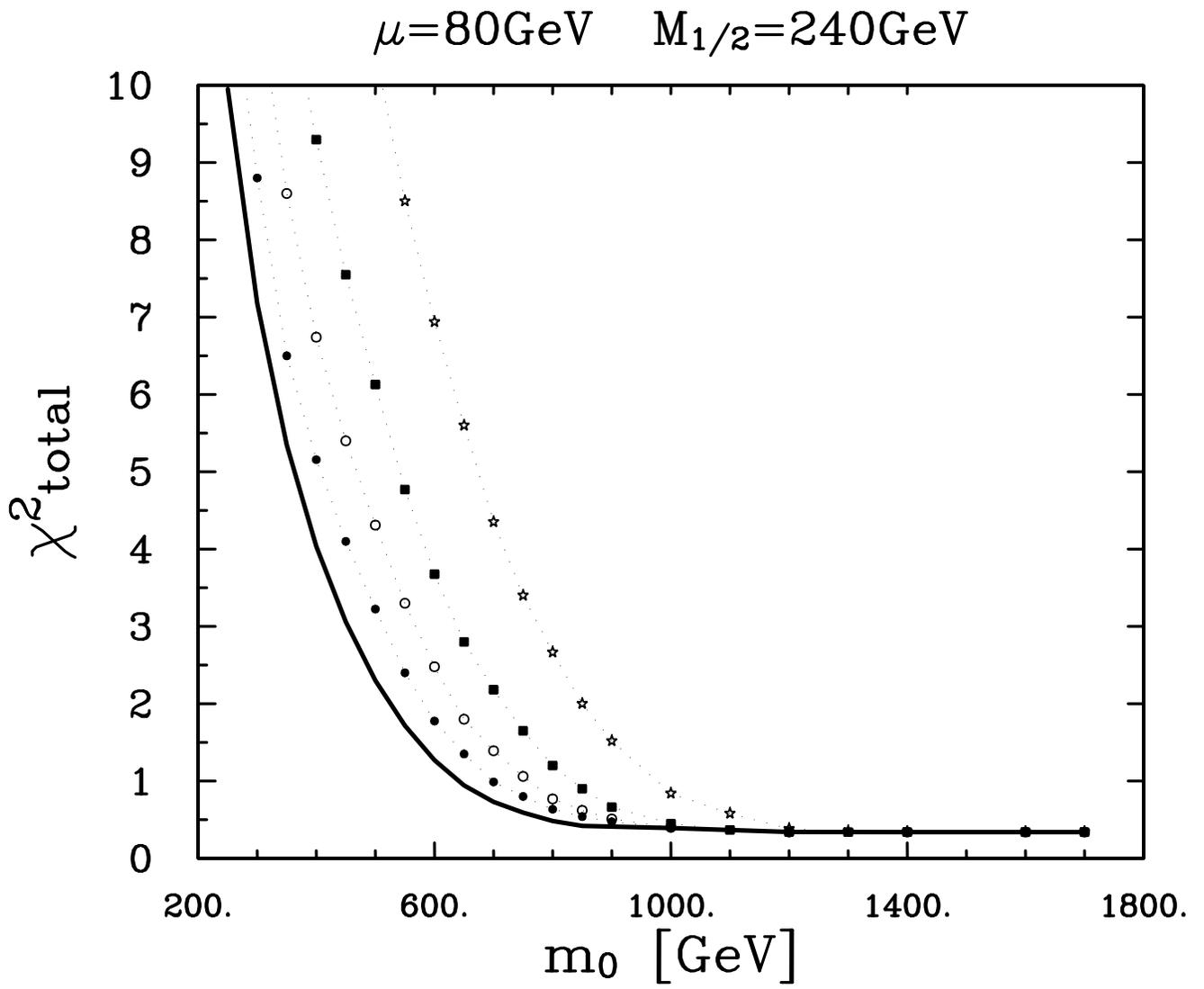
\noindent {\bf Figure 7.}

\noindent 
The variation of the quality of the fit for different lower bounds on 
tree level pseudoscalar mass, in model 4c. The thick solid line represents 
the results of model 4c with $m_A^{tree} > $80GeV. When very close to this 
limit, the pseudoscalar mass receives negative corrections, and its one loop 
corrected value turns out to be at the experimental limit (see tables 
\ref{t:model4cpred} and \ref{t:model4cpredIII} for the results at point 
$m_0$=700GeV). Curves with solid and open circles, solid squares, and stars 
correspond to the lower limit on $m_A^{tree}$ of 130GeV, 200GeV, 300GeV 
and 500GeV, respectively.

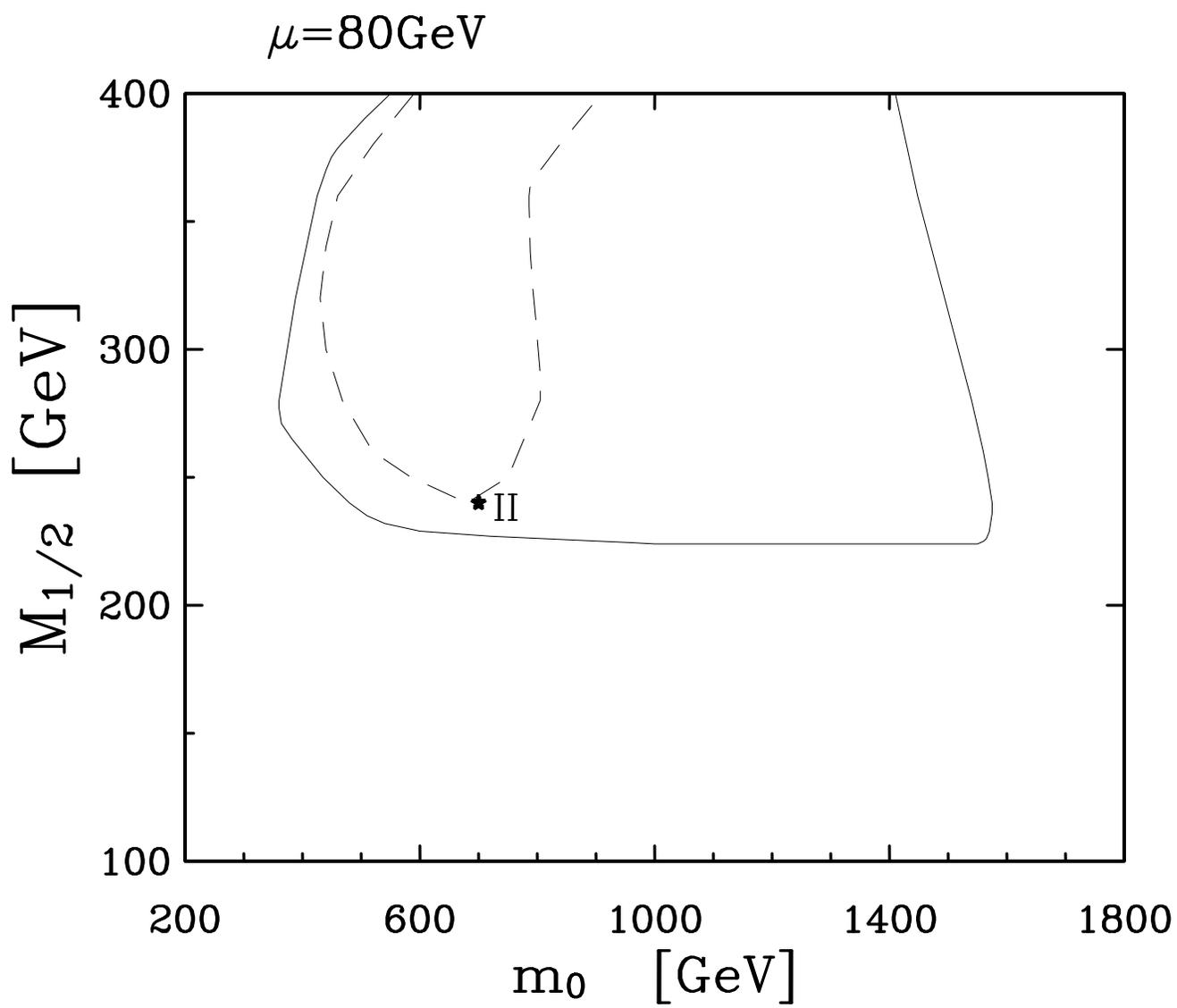
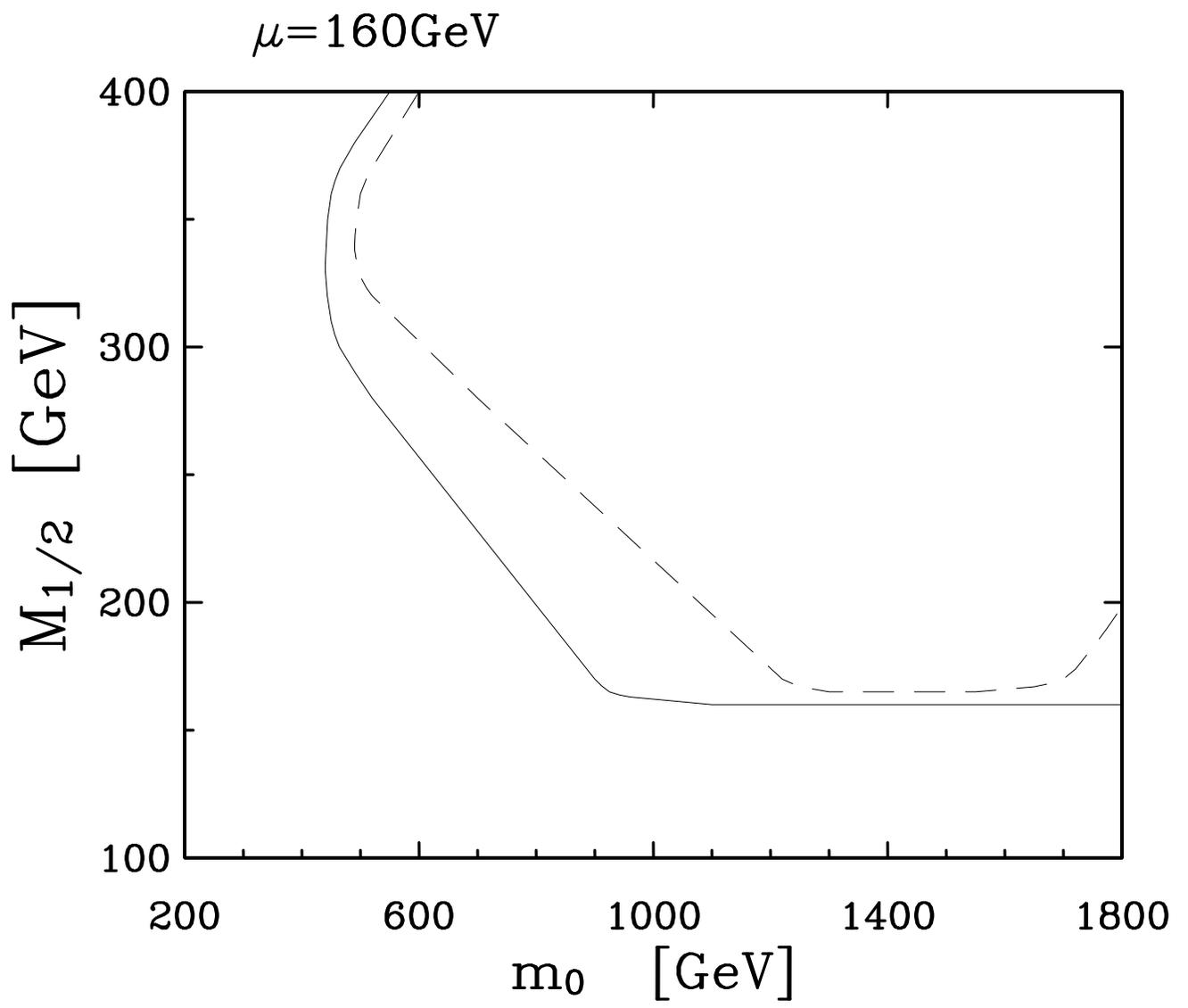
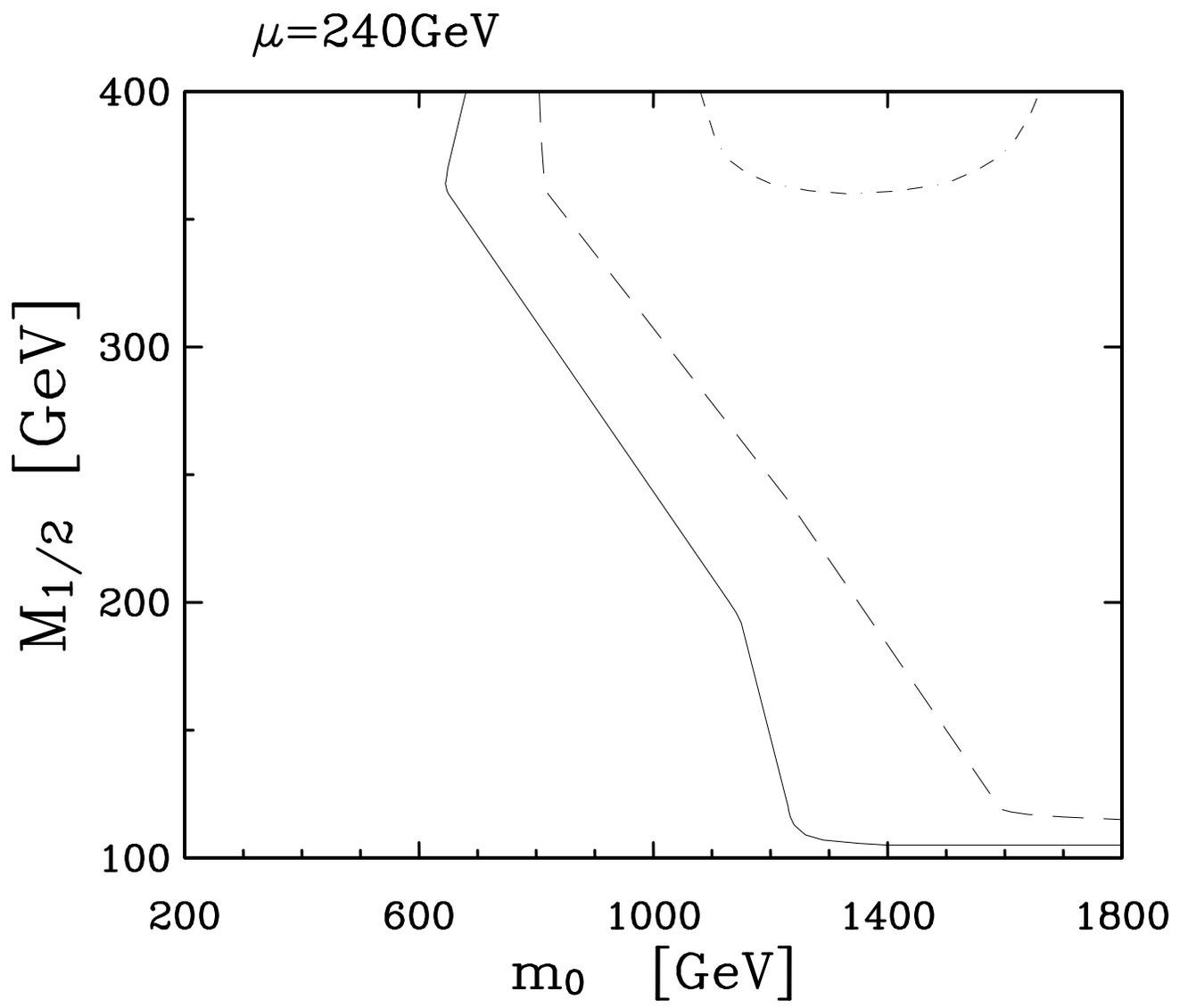
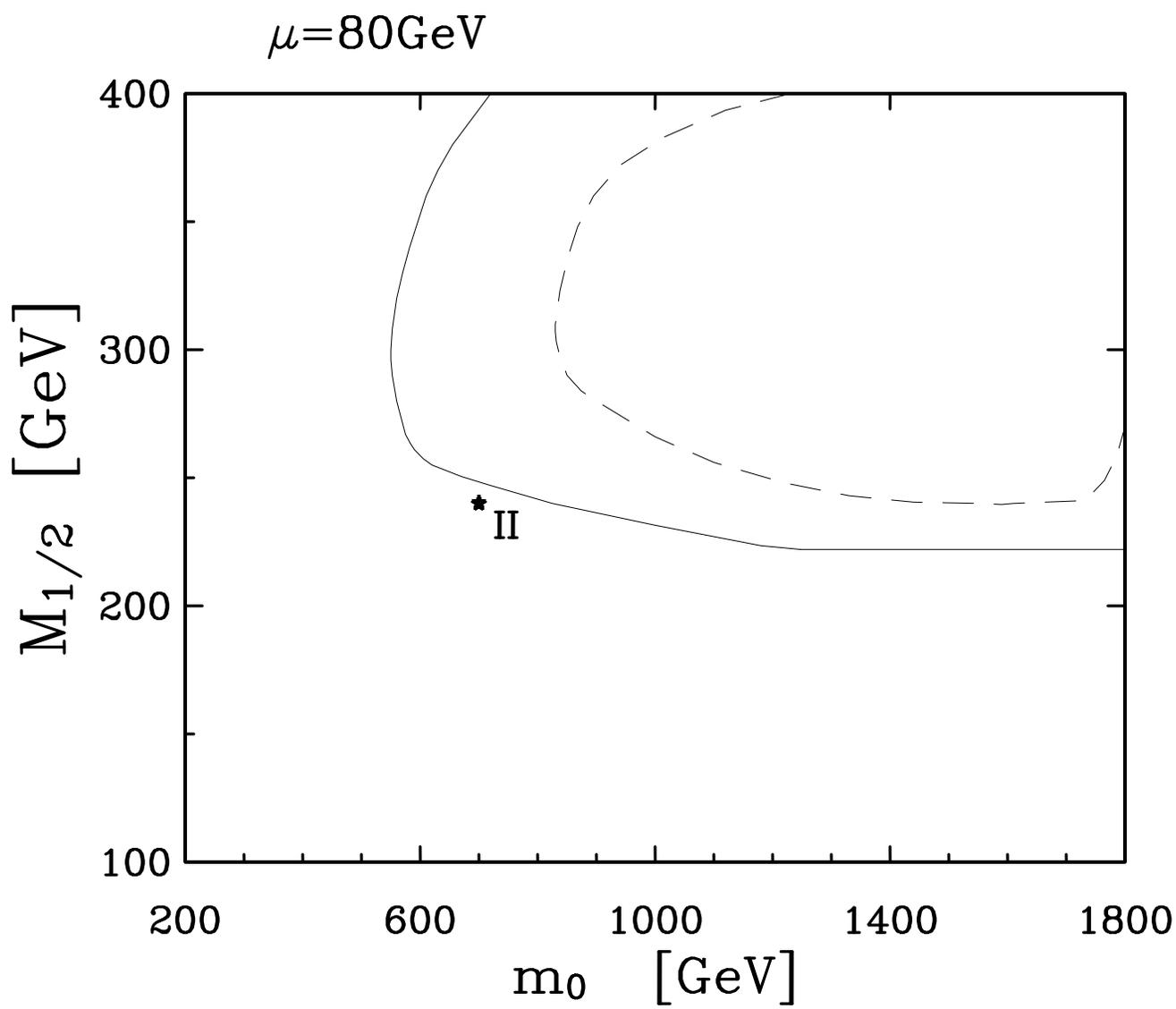
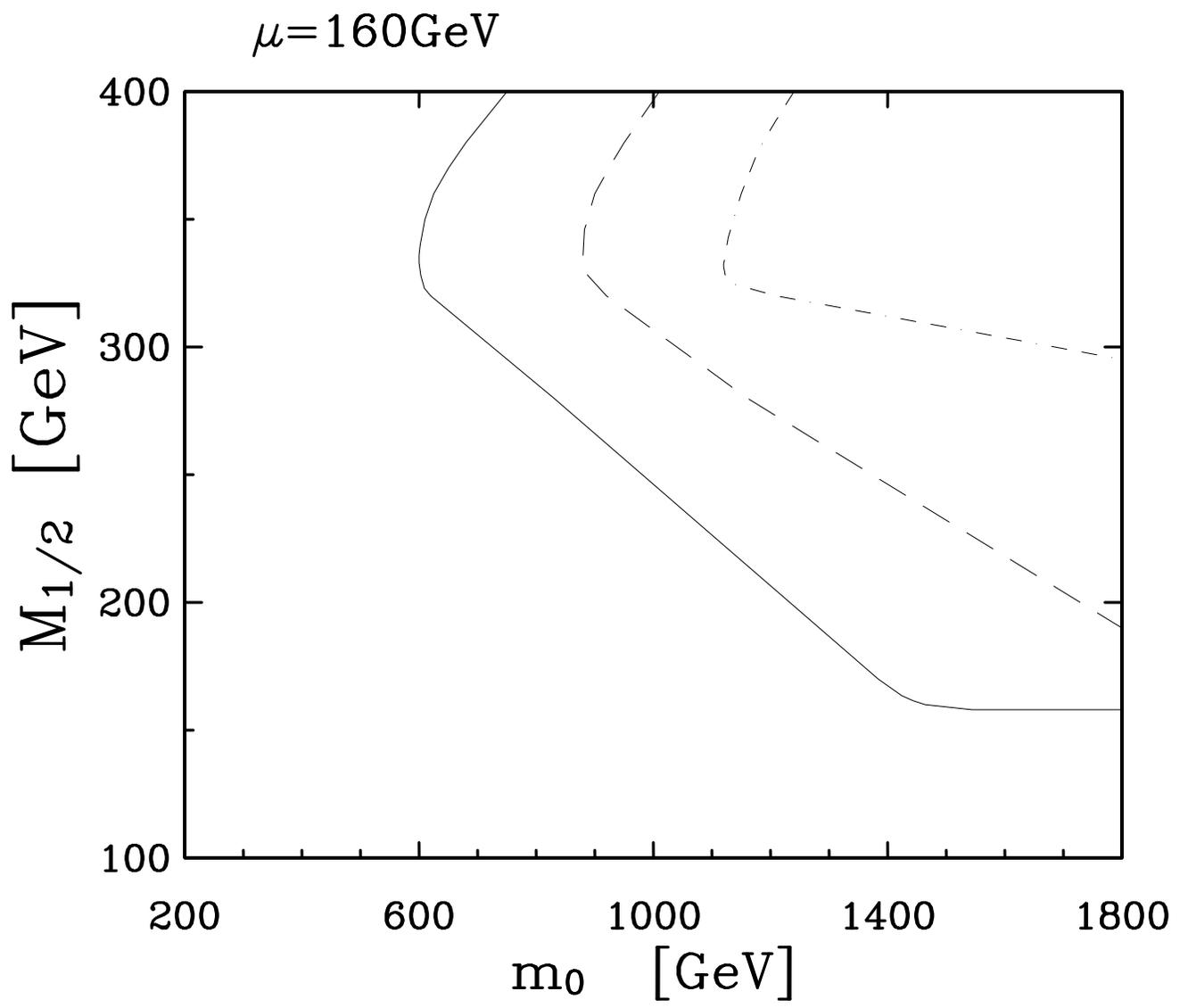
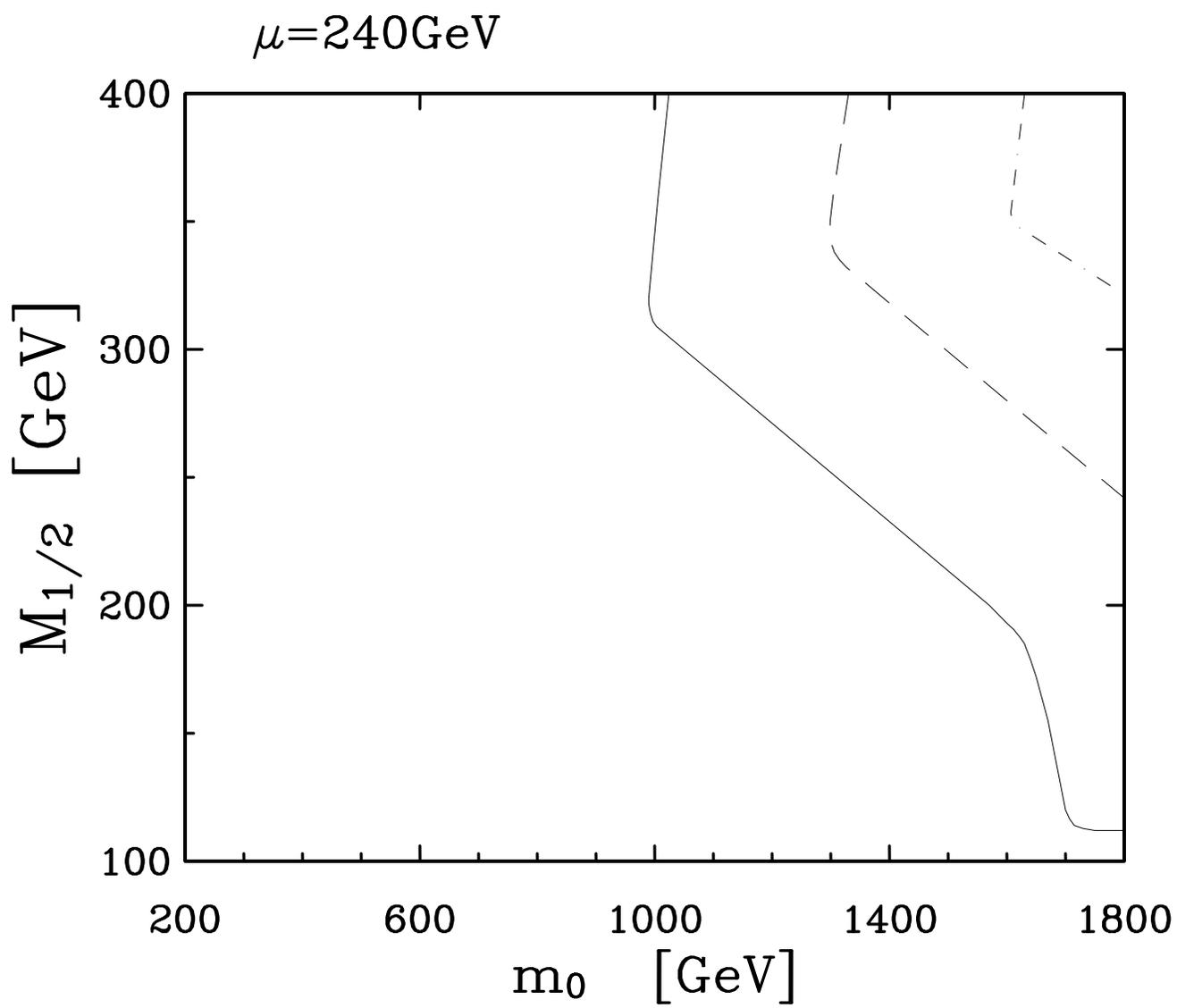
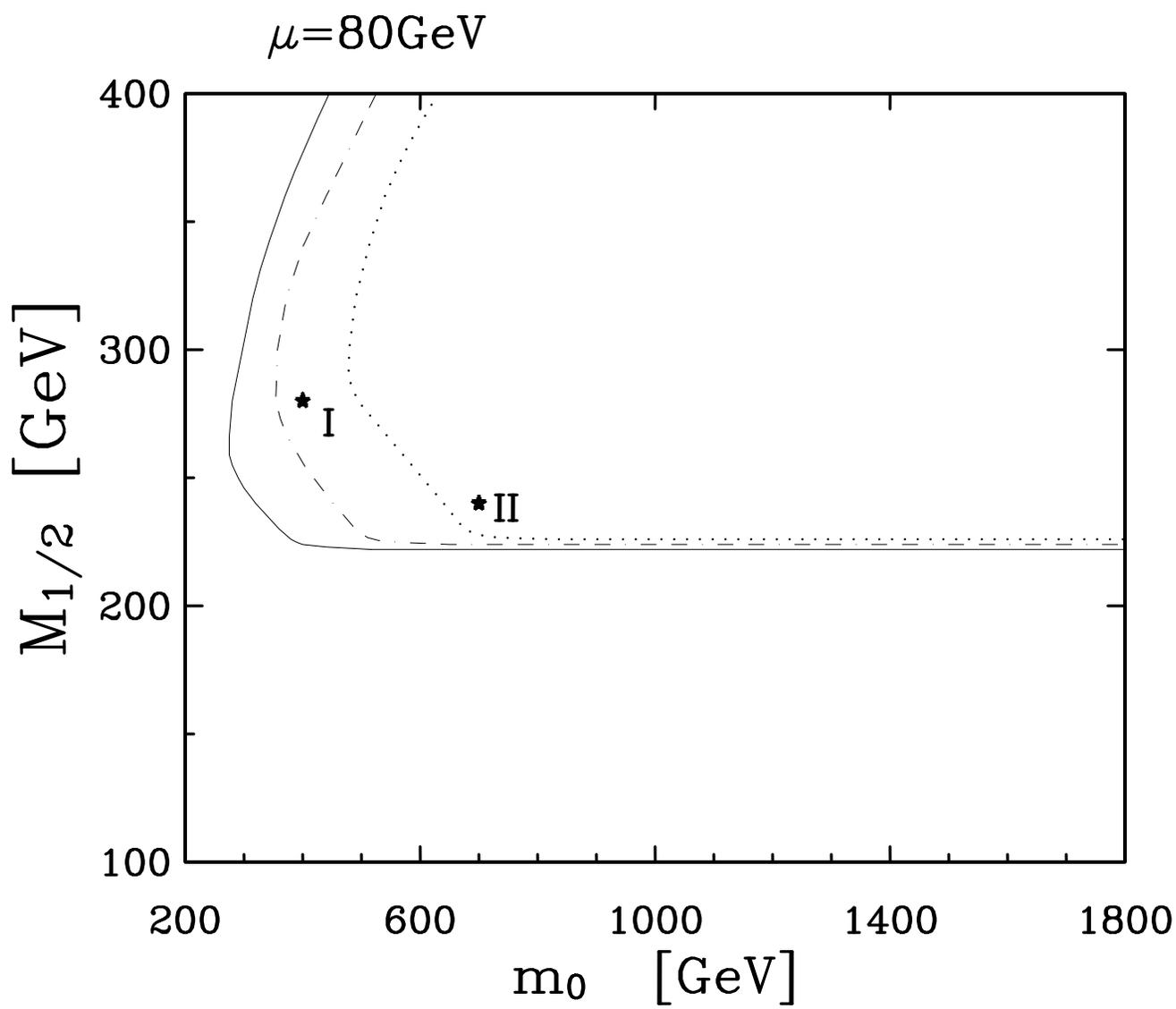
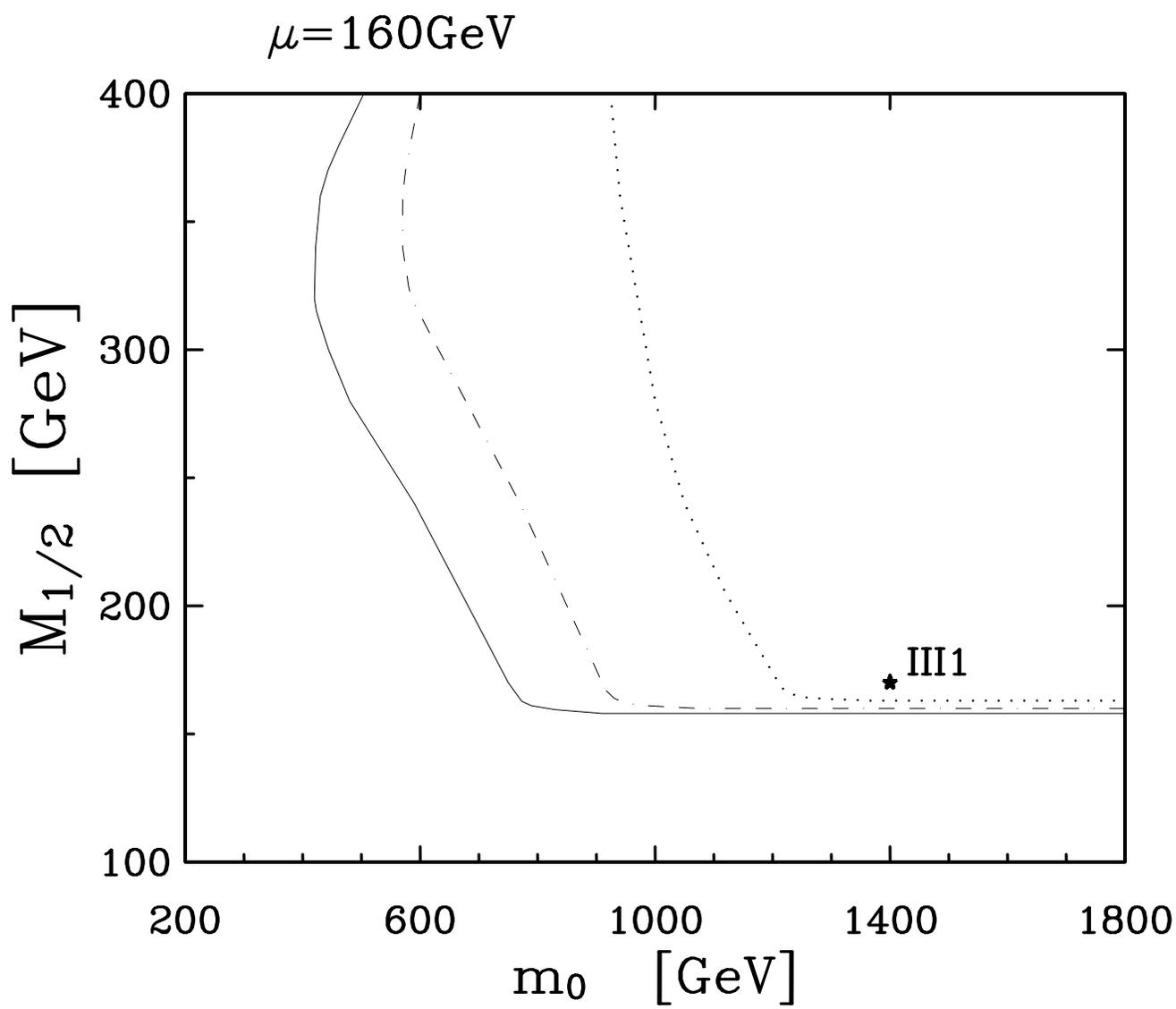
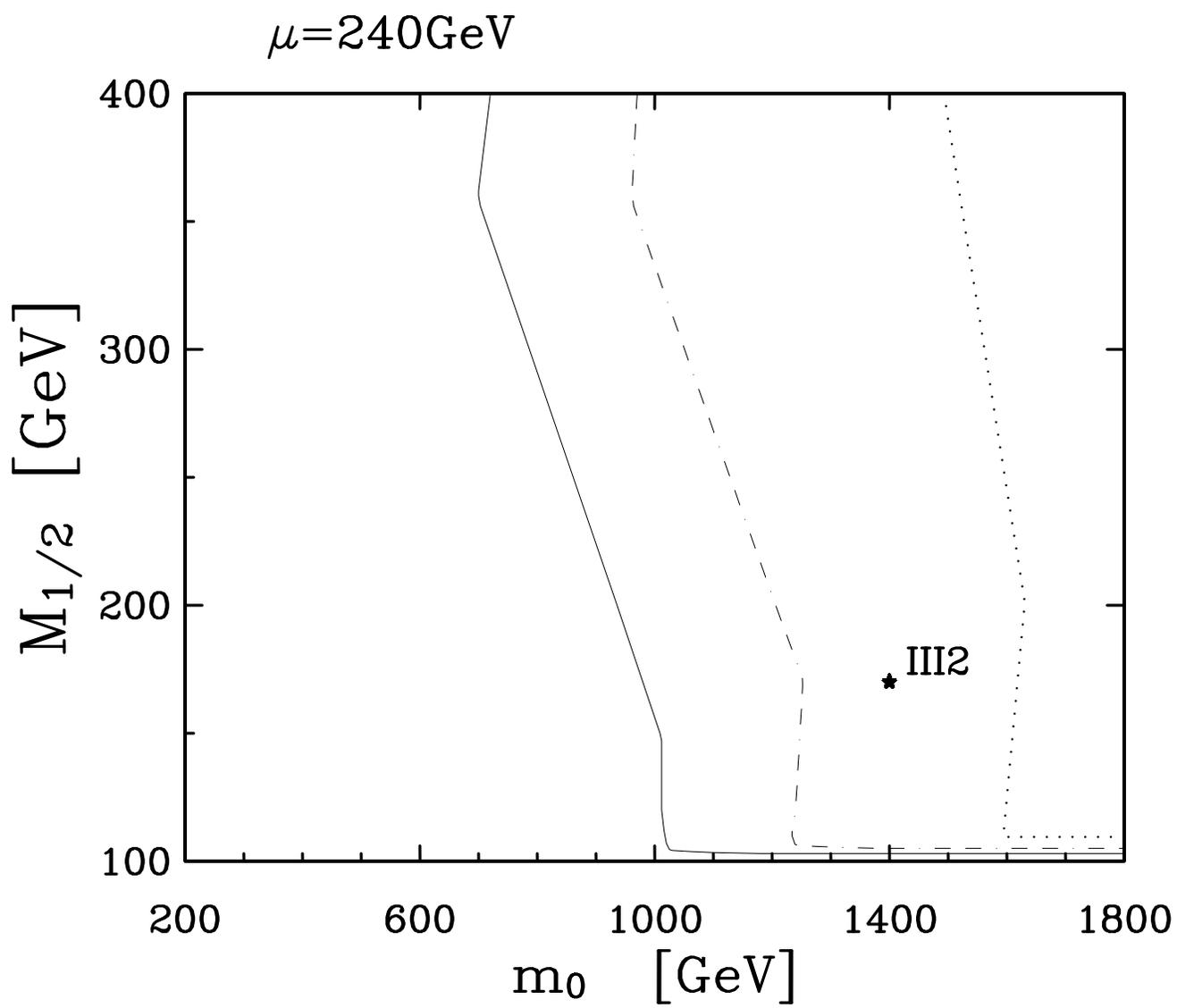
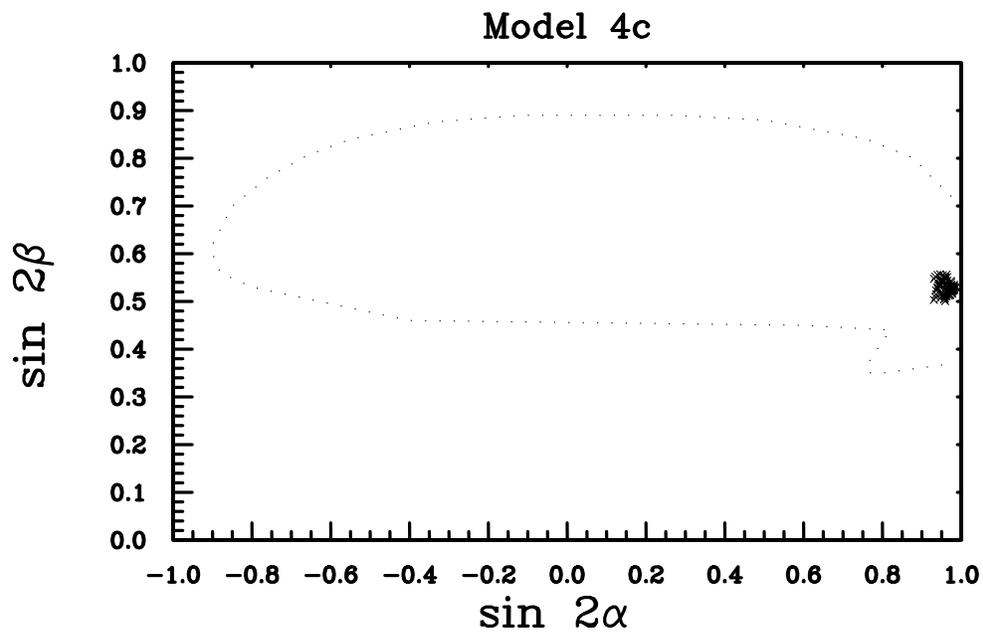
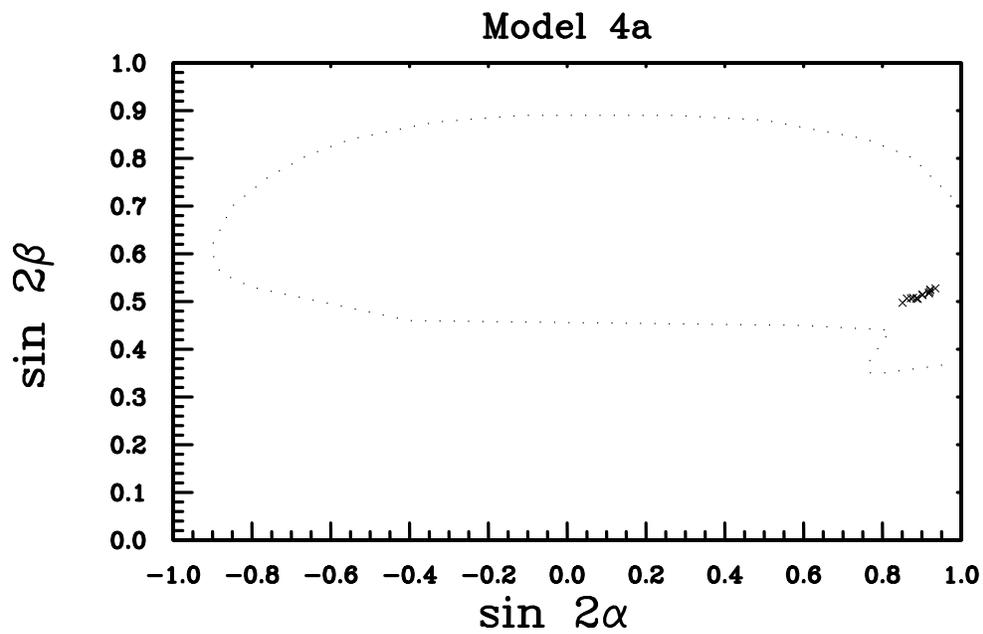
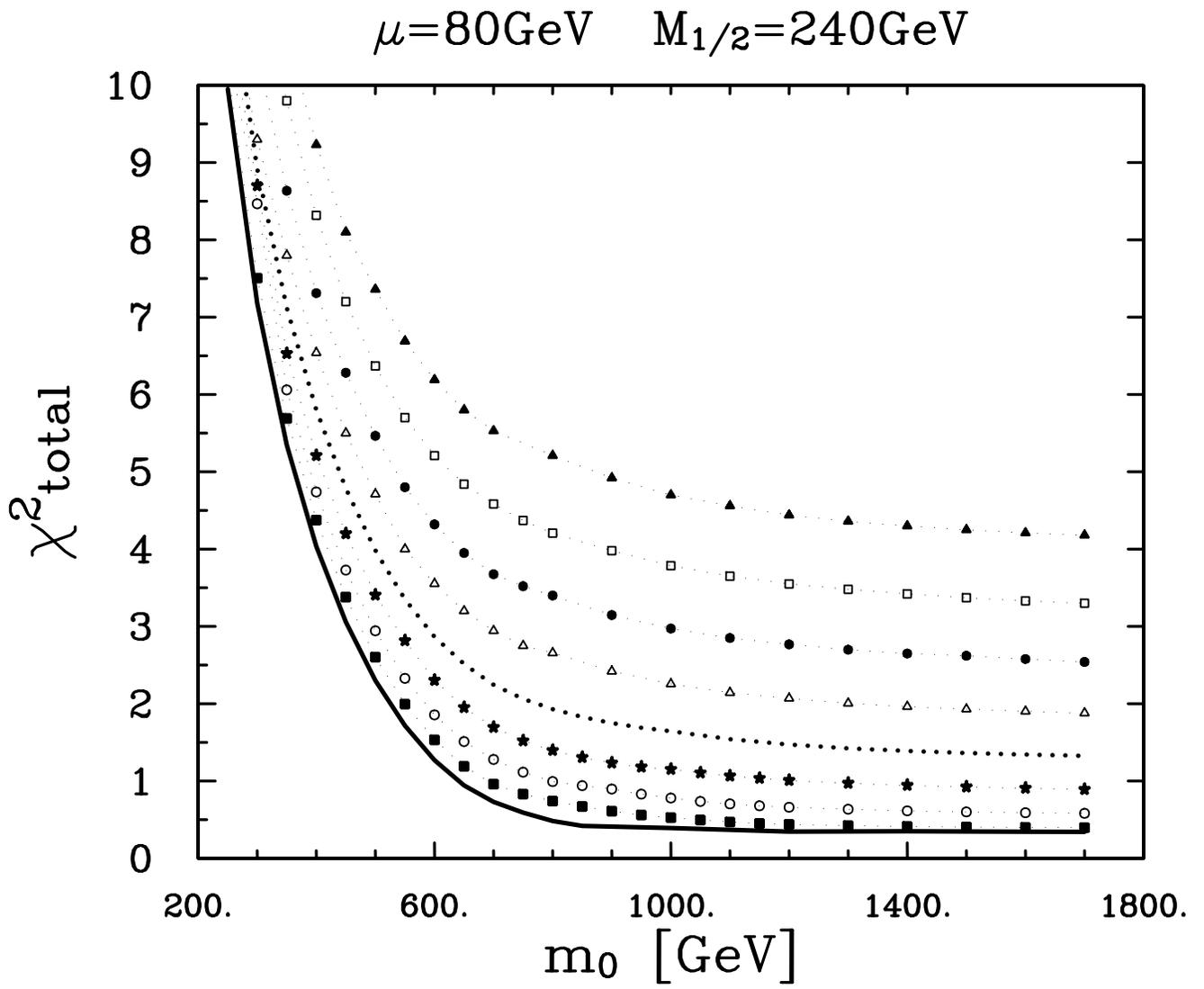
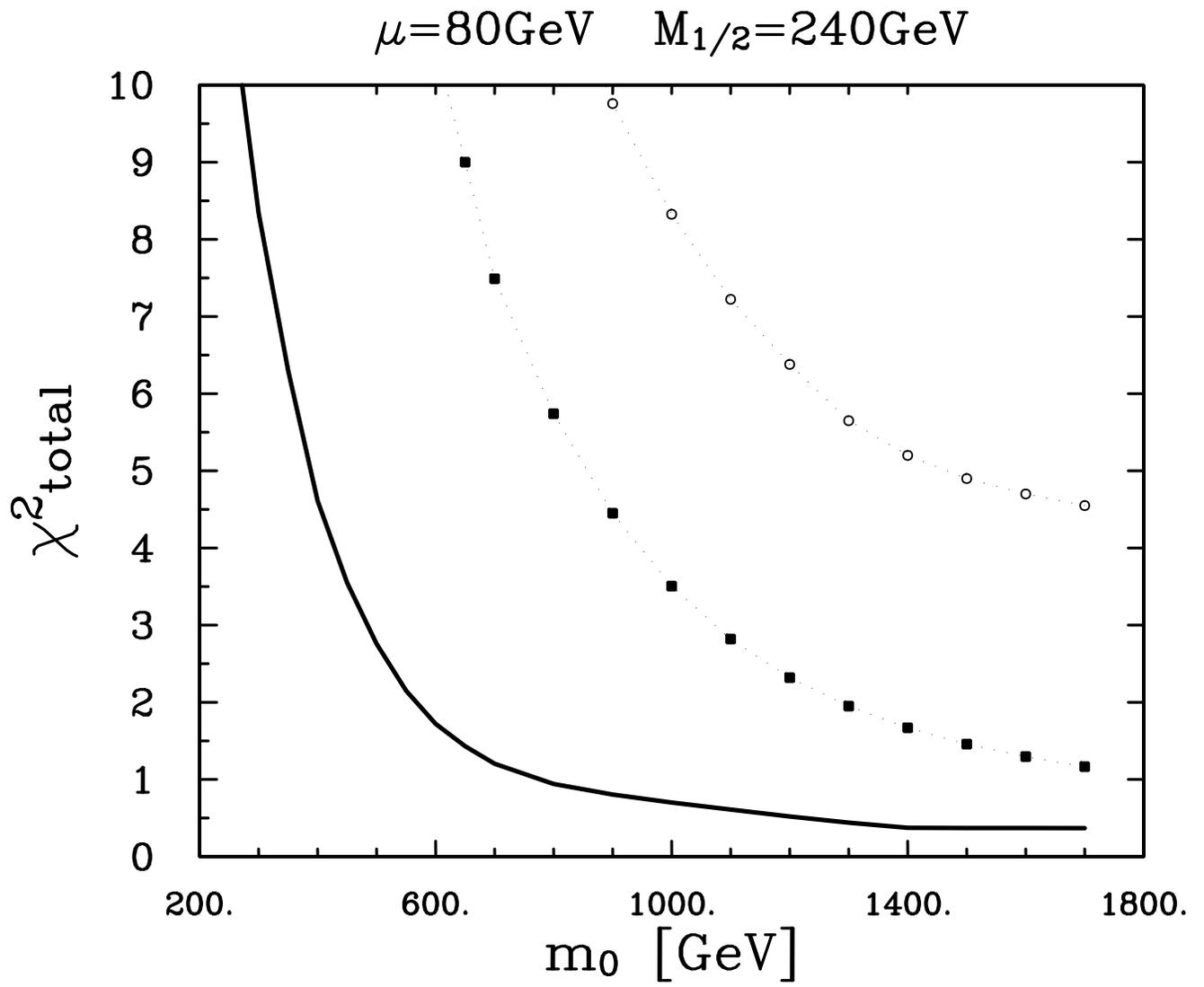
\end{document}